\documentclass[%
 reprint,
bibnotes,
 amsmath,amssymb,
 aps,
]{revtex4-1}

\usepackage{pdfpages}

\usepackage{amsmath, amssymb, verbatim, amsthm, epsfig, psfrag, longtable, subfigure, color}

\usepackage{graphicx}
\usepackage{dcolumn}
\usepackage{bm}

\begin{document}

\preprint{APS/123-QED}

\title{Feedbacks, Receptor Clustering, and Receptor Restriction to Single Cells yield large Turing Spaces for Ligand-receptor based Turing Models}

\author{Tam\'as Kurics$^{1,2,\dagger}$}
\author{Denis Menshykau$^{1,3,\dagger}$}%


\author{Dagmar Iber$^{1,3}$}
 \homepage{http://www.bsse.ethz.ch/cobi/}
\affiliation{
\bf{1} Department for Biosystems Science and Engineering, ETH Zurich, Mattenstrasse 26, 4058 Basel, Switzerland\\
\bf{2} present address:  Institute of Cognitive Neuroscience and Psychology, \\
Research Centre for Natural Sciences, Hungarian Academy of Sciences; Budapest, Hungary\\
\bf{3} Swiss Institute of Bioinformatics (SIB), Switzerland\\
$\dagger$ These authors contributed equally.\\
$\ast$ E-mail: dagmar.iber@bsse.ethz.ch
}%

\date{\today}

\begin{abstract}
\emph{Abstract:} Turing mechanisms can yield a large variety of patterns from noisy, homogenous initial conditions and have been proposed as patterning mechanism for many developmental processes. However, the molecular components that give rise to Turing patterns have remained elusive, and the small size of the parameter space that permits Turing patterns to emerge makes it difficult to explain how Turing patterns could evolve. We have recently shown that Turing patterns can be obtained with a single ligand if the ligand-receptor interaction is taken into account. Here we show that the general properties of ligand-receptor systems result in very large Turing spaces. Thus, the restriction of receptors to single cells, negative feedbacks, regulatory interactions between different ligand-receptor systems, and the clustering of receptors on the cell surface all greatly enlarge the Turing space. We further show that the feedbacks that occur in the FGF10/SHH network that controls lung branching morphogenesis are sufficient to result in large Turing spaces. We conclude that the cellular restriction of receptors provides a mechanism to sufficiently increase the size of the Turing space to make the evolution of Turing patterns likely. Additional feedbacks may then have further enlarged the Turing space. Given their robustness and flexibility, we propose that receptor-ligand based Turing mechanisms present a general mechanism for patterning in biology. \\

\emph{Popular Summary:}
Turing mechanisms can yield a large variety of patterns from noisy, homogenous initial conditions, and have been proposed to underlie many of the patterning phenomena in biology. However, the molecular components are elusive, and the small size of the parameter space that permits Turing patterns makes their evolution unlikely. We show that Turing patterns that arise from ligand-receptor interactions have very large Turing spaces because receptors are restricted to single cells. The Turing space can be further enlarged by additional negative feedbacks, if several receptor-ligand based Turing modules are coupled, and by the clustering of receptors on the cell surface. Given their robustness and flexibility, we propose that receptor-ligand based Turing mechanisms present a general mechanism for patterning in biology.

\begin{description}
\item[PACS numbers] 
\verb 05.65.+b, 87.10.Ca, 87.17.Pq, 87.10.Kn, 87.18.Hf 
\end{description}
\end{abstract}

\maketitle


\section{INTRODUCTION}

\begin{figure*}
\begin{center}
\includegraphics[width=11cm]{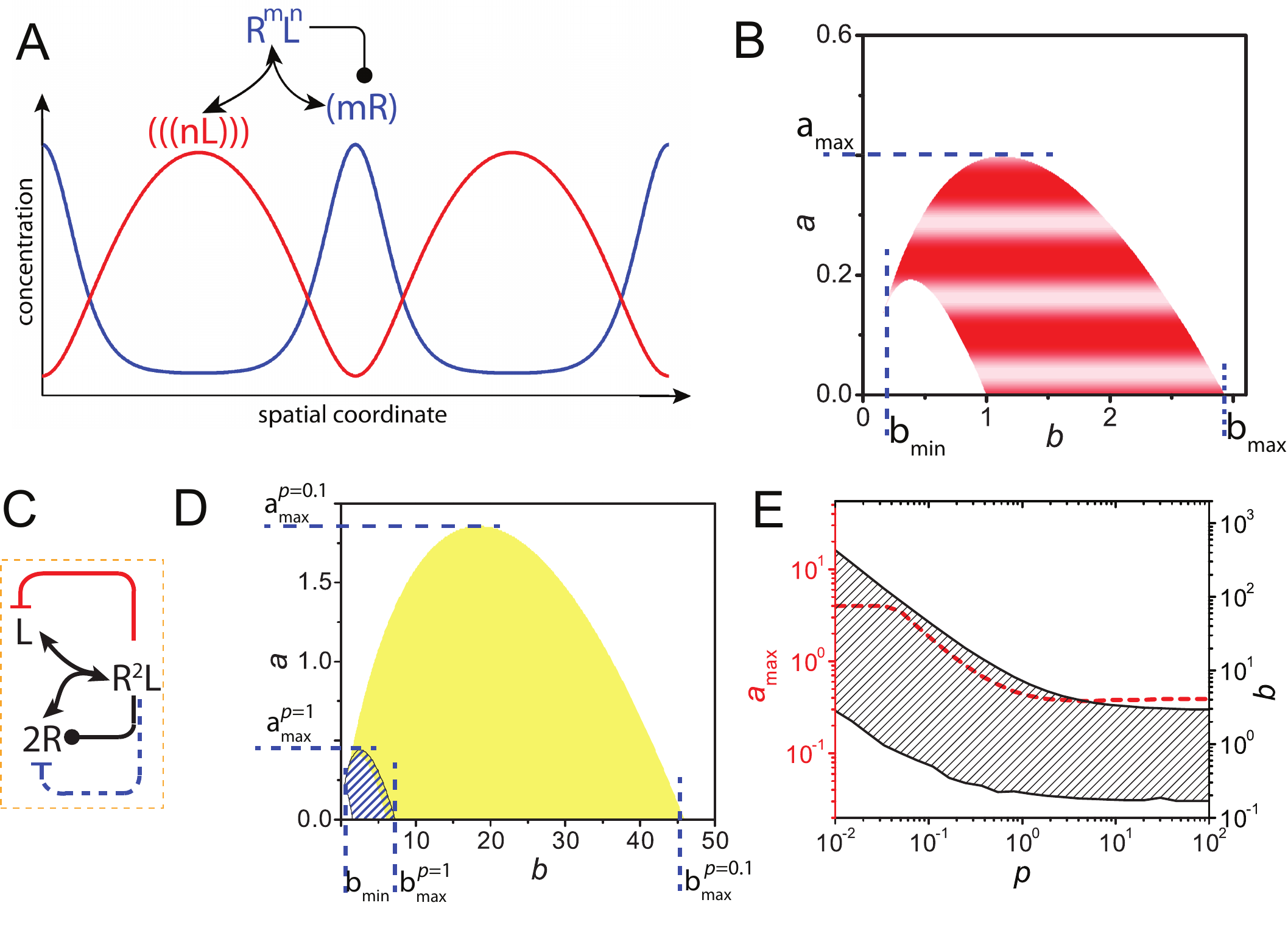}
\end{center}
\caption{
{\bf Ligand-Receptor Interactions can give rise to Turing Patterns.} (Color online) (A) Spatial patterns via a Turing mechanism can result from cooperative receptor-ligand interactions, where $m$ receptors ($R$) and $n$ ligand molecules ($L$) form an active complex that upregulates the receptor concentration by increasing its expression, limiting its turn-over, or similar. Importantly, the highest receptor and ligand concentrations are observed in different places. (B)  In case of the standard network (panel A), Turing patterns emerge only for a small subset of the parameter range of the receptor and ligand production rates, $a$ and $b$. $a_{\textup{max}}$ denotes the maximal value of the receptor production rate, while $b_{\textup{min}}$ and $b_{\textup{max}}$ denote the minimal and maximal ligand production rates. (C) Additional feedbacks (red and dashed, blue arrows) can be mediated by the ligand-receptor complex, $R^2L$; $\leftrightarrow$ indicates receptor-ligand interactions, $\dashv$ inhibitory interactions, and $-\bullet$ up-regulating interactions. (D) The negative feedbacks in panel C (network U5 in Fig. S1) result in a larger Turing space when the response threshold $p$ is lowered from $p=1$ (blue, shaded area) to $p=0.1$ (yellow area). (E) The size of the Turing space for the network in panel C (network U5 in Fig. S1) increases as the response threshold $p$ is lowered. As a measure for the size of the Turing space, we record the maximum of the receptor production rate, $a_{\textup{max}}$, and the ratio of the maximal and minimal ligand production rates $b_{\textup{ratio}} = \frac{b_{\textup{max}}}{b_{\textup{min}}}$, for which Turing patterns can emerge. $a=0$ is part of the Turing space and negative values of $a$ have no physiological interpretation.}
\label{Fig0}
\end{figure*}

The development of complex organisms requires the repeated, reliable emergence of pattern in a cell or tissue from a homogenous, noisy distribution of components, also in the absence of any polarizing queues. It is a long-standing question how stereotyped patterns can emerge during development.  Alan Turing proposed a simple reaction-diffusion-based mechanism \cite{Turing:1952} that has since been shown to have the potential to give rise to a large variety of patterns from noisy, homogenous starting conditions \cite{Kondo:Science:2010, Murray:MathematicalBiology3RdEditionIn2Volumes:2003, Vanag2001}.

Mathematical analysis reveals the types of interactions between the molecular components that can give rise to Turing patterns \cite{Gierer:Kybernetik:1972, Murray:MathematicalBiology3RdEditionIn2Volumes:2003, Prigogine:JChemPhys:1967, Prigogine:TheJournalOfChemicalPhysics:1968}. While many different Turing mechanisms have been proposed to explain pattern formation in biology, it has remained difficult to identify the molecular components \cite{Kondo:Science:2010}. The suggested Turing components are typically two diffusible, extracellular proteins \cite{Cho:Development:2011, Economou:NatGenet:2012, Sick:Science:2006}. However, one of the requirements for Turing patterns is a large difference in the diffusion coefficient between the two Turing components. While a number of chemical systems have been engineered where the diffusion speed of one of the components of the Turing system is strongly reduced, e.g. the Belousov--Zhabotinsky reaction in water-in-oil aerosol microelmulsion \cite{Vanag2001} or in a system with a low-mobility complexing agent \cite{Horvath2009}, these setups do not readily translate to biological systems. For biological systems, it has been suggested that differences in diffusion speed may arise from transient differences in the interactions with the extracellular matrix \cite{Muller:Science:2012}. A number of theoretical studies seek to overcome the requirement of a large difference in diffusivity of Turing components, and an emergence of Turing pattern has been shown to be possible also in the presence of a single diffusive specie coupled to a quenched oscillator \cite{Hsia:PlosComputBiol:2012}; cell migration rather than diffusion has been proposed to result in Turing instabilities \cite{Nakamasu:Pnas:2009, Payne:MolSystBiol:2013}. Finally, cross-diffusion and non-linear diffusion have been shown to support the formation of Turing-type patterns, such that Turing patterns can arise for any ratio of the main diffusivities \cite{Pang2004, Hamidi2012, Butler2011, Ridolfi2011, Fanelli:TheEuropeanPhysicalJournalB:2013, Madzvamuse:JMathBiol:2014}. Cross-diffusion has been shown to arise in crowded environments with finite carrying capacity, i.e. if diffusion is limited when local concentrations or densities reach the carrying capacity \cite{Fanelli:TheEuropeanPhysicalJournalB:2013, Bullara:PhysicalReviewE:2013}. 

Another problem with the applicability of Turing mechanisms to biological pattern formation concerns the size of the parameter space that gives rise to Turing patterns, the Turing space. This parameter space is small for all known Turing mechanisms in the sense that kinetic parameters can be varied only a few fold as long as physiological constraints on the kinetic constants and relative diffusion constants are respected \cite{Murray:JournalOfTheoreticalBiology:1982}. It is therefore unclear how nature could have evolved such mechanism in the first place and how it could have been re-used in different settings during the evolution of new species. Moreover, biological systems are noisy, and time delays as may arise from the multi-step nature of protein expression as well as domain growth and the resulting changes in source and sink terms, may severely affect the existence and type of Turing patterns, though some of these effects as well as further regulatory interactions may somewhat increase the size of the  Turing space \cite{Crampin1999, Gjorgjieva2007, Madzvamuse:JMathBiol:2010, Woolley2011, Maini:Interface:2012, Woolley2012, Klika2012, Gaffney2013}. 

We recently noticed that ligand-receptor interactions of the form shown in Fig.\ \ref{Fig0}A can give rise to Turing patterns \cite{Menshykau:PhysicalBiology:2013, Tanaka:PhysicalBiology:2013, Menshykau:PlosComputBiol:2012, Badugu:SciRep:2012}  as long as the following constraints are met by the the receptor-ligand interaction:
\begin{itemize}
\item Ligands must diffuse much faster than receptors ($d \gg 1$), as is generally the case \cite{Choquet:2003fk,Ries:2009p20248,Kumar2010, Hebert2005}.
\item Receptor-dependent ligand removal must dominate over receptor-independent ligand decay, as is generally the case because unspecific decay is typically much slower than active protein turn-over.
\item Ligands and receptors must bind cooperatively, as is the case for many ligand-receptor pairs \cite{Parkash2008, Jing1996, SpivakKroizman1994, Plotnikov1999, DiGabriele1998, Ibrahimi2005, Goetz2006, Scheufler1999, Nickel2001}.
\item Ligand-receptor complex formation must be fast compared to the other processes, such that we have a quasi-steady state for the ligand-receptor complex concentration. This is the case if the on-rate is very high, i.e.\ binding is diffusion limited, as is the case for many ligand-receptor pairs \cite{ReceptorBook}.
\item The receptor-ligand complex must upregulate the receptor concentration, as has been observed for several receptor systems \cite{Pepicelli1997, Lu2009, Estival1996, Ota2012, Chen1996, Weaver2003, Merino1998}. This positive feedback needs to operate far from saturation, i.e. if we describe the positive regulation by a Hill function of the form $\frac{R^2L}{R^2L + K}$, we require $R^2L \ll K$. Thus, this positive feedback must be rather inefficient.
\end{itemize}
If these conditions are met, the interactions between the receptor, $R$, and the ligand, $L$, result in Schnakenberg-type kinetics \cite{Schnakenberg:JTheorBiol:1979} of the form
\begin{eqnarray}
\hspace*{-0.4cm}  \frac{\partial R}{\partial t} &=& \Delta R+ \gamma f(R,L)  	\  \textrm{with} \	 f(R,L)=a-R+R^2 L \label{eq:1} \\
\hspace*{-0.4cm}  \frac{\partial L}{\partial t} &=& d \Delta L+ \gamma g(R,L)	\  \textrm{with} \   g(R,L)=b-R^2 L,  \label{eq:2}
\end{eqnarray}
which correspond to the so-called activator-depleted substrate Turing kinetics, first described by Gierer and Meinhardt \cite{Gierer:Kybernetik:1972}, and which are very similar to the chemical Turing system first described by Prigogine and co-workers \cite{Prigogine:TheJournalOfChemicalPhysics:1968}.  The detailed derivation of these equations for receptor-ligand interactions can be found in previous publications  \cite{Menshykau:PhysicalBiology:2013, Tanaka:PhysicalBiology:2013, Menshykau:PlosComputBiol:2012, Badugu:SciRep:2012} and in the Appendix I. The $\Delta R$ and $d\Delta L$ terms represent the diffusion terms, where $d$ is the relative diffusion constant of ligand and receptor. Ligands typically diffuse faster than their receptors, $d \gg1$ \cite{Choquet:NatRevNeurosci:2003, Kumar:BiophysJ:2010, Ries:NatMethods:2009}, thus naturally meeting the Turing condition of different diffusivities. Receptor diffusion is restricted to single cells, and we have previously shown that patterns also emerge on such cellularized domains \cite{Menshykau:PlosComputBiol:2012}. The constants $a$  and $b$  are the receptor and ligand production rates. The $-R$ term describes the ligand-independent decay of the receptor at a rate proportional to the available receptor concentration, so-called linear decay. The term $R^2 L$ represents the quasi-steady state concentration of the receptor-ligand complex. Signaling complexes with a different stochiometry  also result in Turing patterns \cite{Menshykau:PlosComputBiol:2012}. The 'minus' term in Eq.\ \eqref{eq:2} then reflects the receptor-dependent ligand removal rate, while the 'plus' term in Eq.\ \eqref{eq:1} reflects the combined effects of ligand-induced receptor removal and ligand-induced receptor accumulation on the cell membrane (by increased transcription, translation, recycling, less constitutive removal or similar). The $\gamma$ term arises in the non-dimensionalization of the model (Eq. \ref{eq:Schnakenberg}) and is useful as it is proportional to the domain area, and it gives the relative strength of the reaction and diffusion terms \cite{Murray:MathematicalBiology3RdEditionIn2Volumes:2003}.

A number of ligand-receptor systems meet the above conditions, including Hedgehog and its receptor PTCH \cite{Tanaka:PhysicalBiology:2013, Menshykau:PlosComputBiol:2012, Chen1996, Weaver2003, Goetz2006}, BMPs and their BMP receptors \cite{Badugu:SciRep:2012, Scheufler1999, Nickel2001, Merino1998}, GDNF and its receptor RET \cite{Menshykau:PhysicalBiology:2013, Parkash2008, Jing1996, Pepicelli1997, Lu2009}, as well as FGFs and their FGF receptors \cite{SpivakKroizman1994, Plotnikov1999, DiGabriele1998, Ibrahimi2005, Estival1996, Ota2012}. Thus, all of these proteins are multimers, and, by a range of mechanisms, the formation of the multimeric ligand-receptor complexes enhances the concentration of receptors on the membrane, as recently reviewed \cite{Iber:OpenBiol:2013}. We further showed that models based on these proteins could recapitulate the relevant wildtype and mutant expression patterns in the respective developmental systems \cite{Menshykau:PhysicalBiology:2013, Tanaka:PhysicalBiology:2013, Menshykau:PlosComputBiol:2012, Badugu:SciRep:2012, Celliere:BiologyOpen:2012}.

Here we show that ligand-receptor based Turing mechanisms can have significantly enlarged Turing spaces if we include negative feedbacks or couple several Turing modules, as generally found in biological systems. Similarly, the restriction of receptors to single cells and their clustering further increases the size of the Turing space. We conclude that a receptor-ligand based Turing mechanism offers a realistic mechanism to implement the Turing mechanism in a biological setting. The observation that the restriction of receptors to cells is sufficient to massively increase the Turing space offers an explanation of how Turing patterns may have first evolved in nature; additional feedbacks could then further enlarge the Turing space.

\section{RESULTS}

 The Turing mechanism has been analysed extensively, and the parameter space that permits Turing patterns to emerge can easiest be determined with the help of a linear stability analysis \cite{Murray:MathematicalBiology3RdEditionIn2Volumes:2003}; see the Appendix II. To keep the analysis feasible, it is advisable to consider as simple models as possible, and to restrict the number of parameters to a minimum. The non-dimensional ligand-receptor based Turing model (Eq.\ \eqref{eq:1}-\eqref{eq:2}) has four parameters: the relative ligand/receptor diffusion constant $d$, the receptor production rate $a$, the ligand production rate $b$, and the scaling factor $\gamma$. The parameters $a$, $b$, $d$ determine whether Turing patterns can emerge, while the scaling factor $\gamma$ determines whether the domain is sufficiently large for Turing patterns to emerge. We therefore do not need to analyse $\gamma$ here. The relative diffusion constant of ligands and receptors, $d$, affects the size of the Turing space in that a larger $d$ results in a larger Turing space \cite{Murray:MathematicalBiology3RdEditionIn2Volumes:2003}. Since this effect is well documented, but limited by the physiological difference between the diffusion constants of ligands and receptors we fixed the relative diffusion constant in our analysis. For a simple receptor-ligand based Turing system, in which receptor and ligand bind cooperatively and upregulate the receptor concentration (Fig.\ \ref{Fig0}A), both parameter values $a$ and $b$ produce Turing patterns only within a small range (Fig.\ \ref{Fig0}B), i.e.\ the ligand production rate can at most be halved or doubled without leaving the Turing space. The Turing space is thus very small, even though the relative diffusion constant, $d=50$, between ligands and receptors was chosen to be rather large compared to what could be justified for two soluble ligands. We will now analyse the impact of feedbacks, receptor clustering, and the restriction of receptors to single cells on the size of the Turing Space.

\begin{figure}
\begin{center}
\includegraphics[width=\columnwidth]{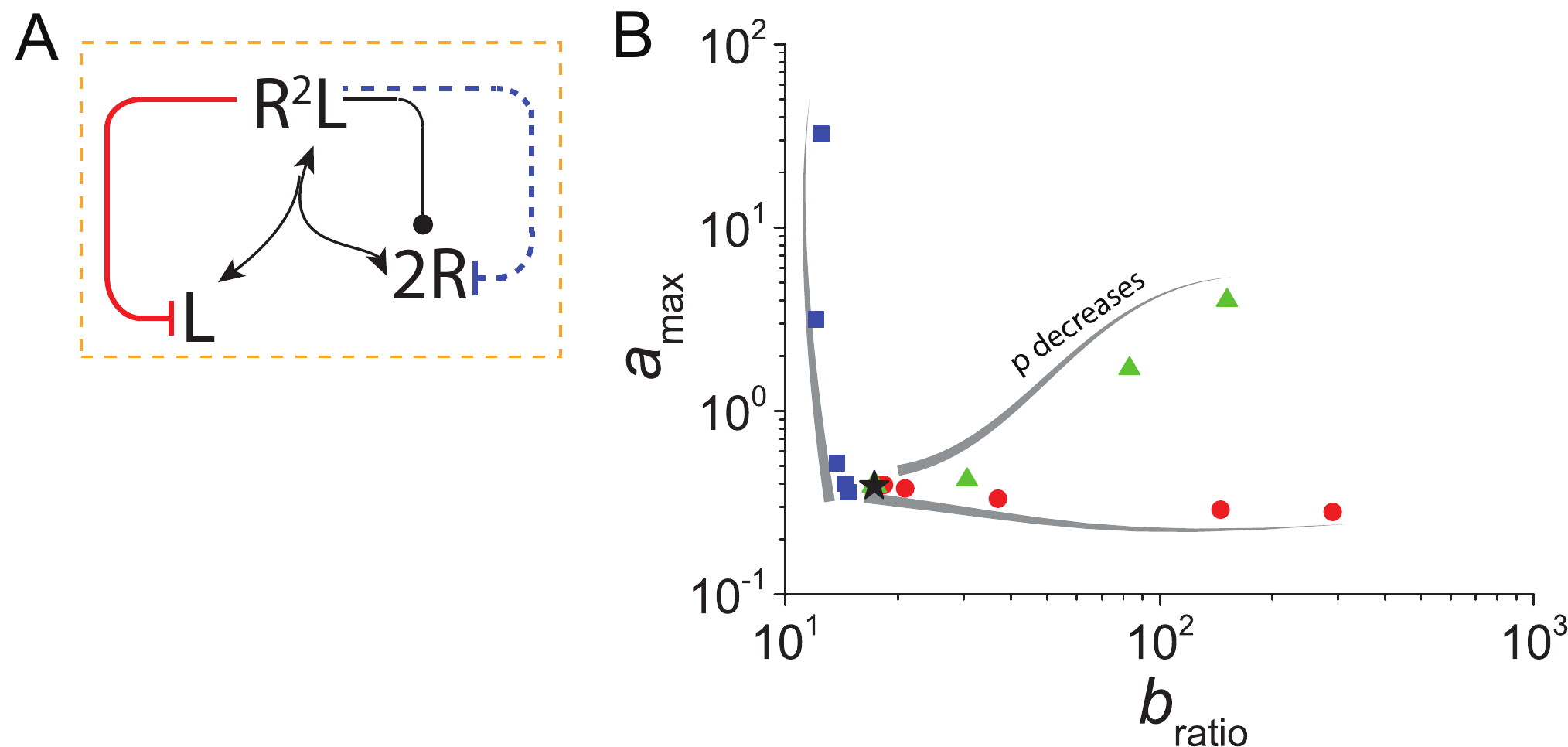}
\end{center}
\caption{
{\bf Negative Feedbacks by Receptor-Ligand Complexes result in Turing patterns with large Turing Spaces.} (Color online) (A) The simulated network architecture. Two receptors $R$ interact with one dimeric ligand $L$ to form a receptor-ligand complex $R^2L$ (black arrows, $\leftrightarrow$). The receptor-ligand complex upregulates the presence of receptor ($-\bullet$). In addition to these core interactions that can result in a Turing mechanism, we considered negative feedbacks ($\dashv$) on the ligand production (red arrow) and / or the receptor production (blue, dashed arrow).  (B) A negative feedback on the receptor production rate (blue dashed arrow in panel A) increases the Turing parameter space for the receptor production rate, $a$ (blue squares in panel B) compared to the standard network (black part of the network in panel A and black star in panel B). A negative feedback on ligand production (red, solid arrow in panel A) enlarges the Turing parameter space for the ligand production rate, $b$, (red circles in B). In the presence of both feedbacks the Turing parameter space is enlarged along both axes  (green triangles in panel B). The feedback effects are stronger, the lower the feedback threshold, $p$ ($p$= 0.01, 0.1, 1, 10, 100). The grey arrow indicates the direction, in which the feedback threshold, $p$ decreases.}
\label{Fig1u}
\end{figure}

\subsection{The impact of Feedbacks on the Turing Space of a single receptor-ligand based Turing Module}

Feedbacks are ubiquitous in biological signaling systems. In the framework of receptor-ligand-based Turing mechanisms, feedbacks result from regulatory interactions of the receptor-ligand complex, $R^2L$ (Fig.\ \ref{Fig0}C). To encode feedbacks mediated by receptor-ligand signaling, we modified the reaction terms $f(R,L)$ and $g(R,L)$ in the Turing model (Eq.\ \eqref{eq:1}-\eqref{eq:2}). See Supplemental Material \cite{Supp} for the list of all tested models with additional feedbacks. Thus a positive feedback on receptor or ligand expression would be obtained by adding a term $p R^2L$ to the respective equation and/or by multiplying the constitutive receptor and ligand production rates $a$ and $b$ with the factor $\frac{R^2 L}{R^2 L+p}$. A negative feedback would be obtained by multiplying the constitutive receptor and ligand expression rates $a$ and $b$ with the factor $\frac{1}{R^2 L / p+1}$. The new parameter $p$ represents the response threshold to the receptor-ligand complex. Figure \ref{Fig0}D illustrates the impact of feedbacks on the Turing space for the regulatory system with two additional negative feedbacks shown in Fig.\ \ref{Fig0}C. For a large response threshold ($p=1$) the Turing space is similar in size to the non-feedback case (compare the blue shaded area in Fig.\ \ref{Fig0}D to the Turing space in Fig.\ \ref{Fig0}B). As we lower the response threshold to $p=0.1$ and thus increase the strength of the negative feedbacks the Turing space increases in size, i.e.\ both the maximal receptor production rate, $a_{\textup{max}}$, as well as the range of ligand expression rates $[b_{\textup{min}}, b_{\textup{max}}]$ increases (yellow area in Fig.\ \ref{Fig0}D); the minimum of $a$ is negative and $a_{\textup{max}}$ thus defines  the size of the physiological parameter range, $[0, a_{\textup{max}} ]$. As the response threshold $p$ is lowered further, the size of the Turing space further increases (Fig.\ \ref{Fig0}E). 

We next systematically analysed eleven positive, negative, and mixed feedback architectures that were obtained by including feedbacks of the receptor-ligand complex ($R^2L$) on the receptor ($a$) and/or ligand production rates ($b$), as well as on the rate of receptor up-regulation upon receptor-ligand binding (for details see Appendix II, Fig.\ S1). Figure \ref{Fig1u}A,B shows the three cases with the largest Turing space out of the eleven cases analyzed. For better readability, we only record the maximal receptor production rate, $a_{\textup{max}}$ as well as the ratio, $b_{\textup{ratio}} = \frac{b_{\textup{max}}}{b_{\textup{min}}}$, of the maximal and minimal ligand production rates that permit Turing patterns to emerge. We note that the ratio $b_{\textup{ratio}} = \frac{b_{\textup{max}}}{b_{\textup{min}}}$ is biologically more relevant than the absolute size of the Turing space, $\Delta b= b_{\textup{max}} - b_{\textup{min}}$, because in biology relative changes in regulatory control and thus in production rates are particularly relevant; the absolute values are typically very difficult to measure. The largest Turing spaces are obtained with negative feedbacks. When the negative feedback is applied to the constitutive receptor expression, $a$ (blue), the maximal value of $a$ increases relative to the standard model (black spot) as the response threshold, $p$, is lowered; the minimum of $a$ is negative and $a_{\textup{max}}$ thus defines the size of the physiological parameter range, $[0, a_{\textup{max}}]$. If a feedback is applied to the ligand expression rate, $b$, then, as the response threshold, $p$, is lowered, the range of $b$ increases (red) compared to the standard model (black spot). The largest Turing spaces, expanded both along the $a$ and $b$ axes, are observed when negative feedbacks are applied to both, the receptor and ligand expression rates (green). The impact of the negative feedbacks can be observed for a wide range of the new parameters, $p$, and becomes stronger the smaller the value of the response threshold $p$ (Fig.\ \ref{Fig1u}B). As the response threshold $p$ is increased, the maximal values of $a$, and the range of $b$ all attain the value of the standard receptor-ligand model and thus all converge in the black spot in Fig.\ \ref{Fig1u}B. In summary, substantially enlarged Turing spaces are observed when signaling by the the receptor-ligand complex lowers the receptor production rate (Fig.\ \ref{Fig1u}B, blue), or the ligand production rate (Fig.\ \ref{Fig1u}B, red), or both (Fig.\ \ref{Fig1u}B, green).

\begin{figure}
\begin{center}
\includegraphics[width=\columnwidth]{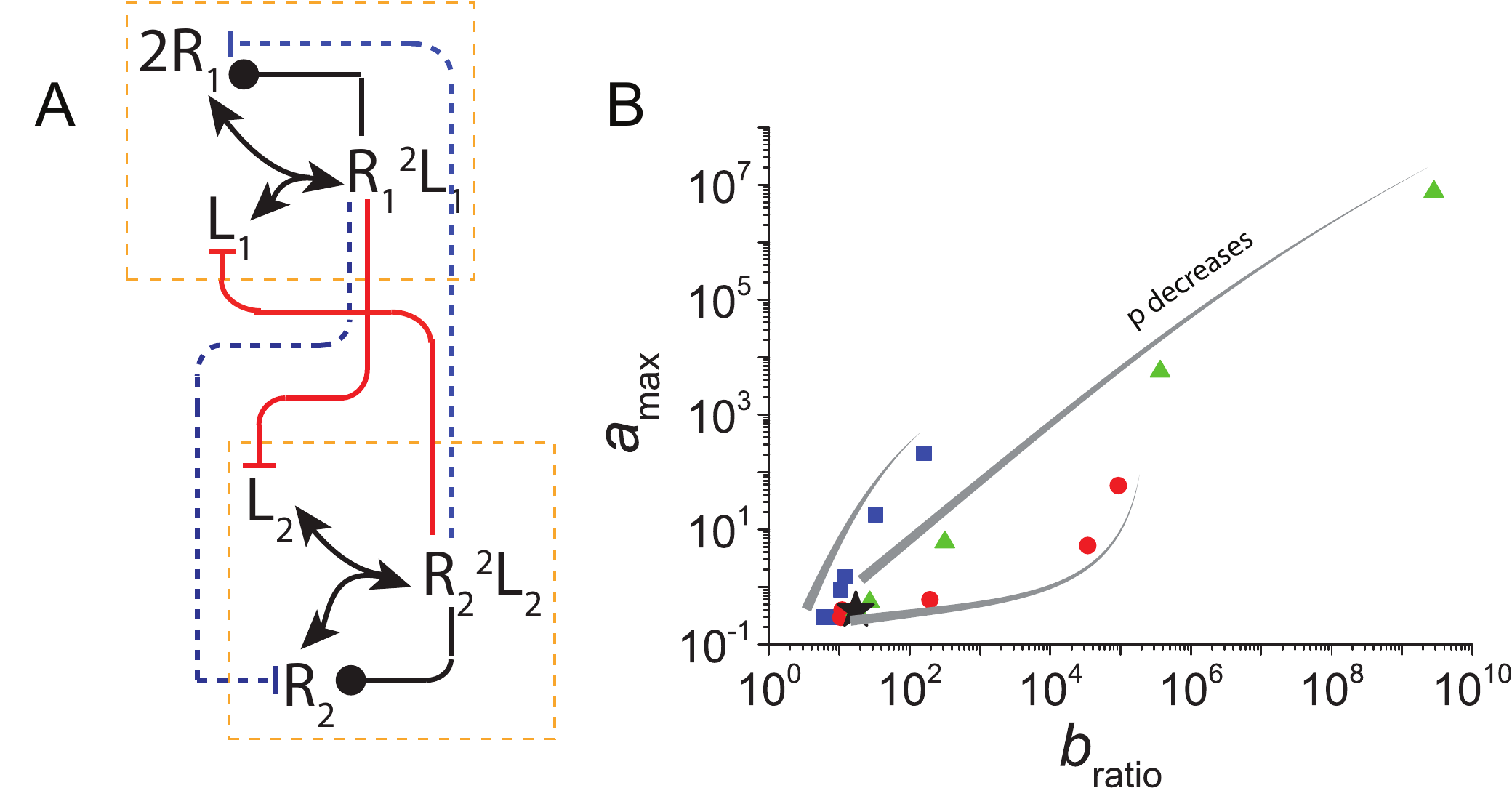}
\end{center}
\caption{
{\bf Coupling of several Receptor-Ligand based Turing modules further enlarges the Turing Space.} (Color online) (A) The simulated network architecture. Two receptor-ligand based Turing modules, as analysed in Fig. 2 (black arrows, $\leftrightarrow$, $-\bullet$), are coupled via  additional negative feedbacks ($\dashv$) on the ligand production rates (red, solid arrows) and / or the receptor production rates (blue, dashed arrows).  (B) A negative feedback on the receptor production rate (dashed, blue line in panel A) increases the Turing parameter space for the receptor production rate, $a$ (blue squares in panel B) compared to the standard network (black part of the network in Fig. 2A and black star in panel B).  A negative feedback on ligand production (red, solid arrow in panel A) enlarges the Turing parameter space for the ligand production rate, $b$, (red circles in B). In the presence of both feedbacks the Turing parameter space is enlarged along both axes  (green triangles in panel B). The feedback effects are stronger, the lower the feedback threshold, $p$ ($p$= 0.01, 0.1, 1, 10, 100). The grey arrow indicates the direction, in which the feedback threshold, $p$ decreases.}
\label{Fig1c}
\end{figure}

\subsection{Coupled Turing Modules}

In patterning processes, several receptor-ligand systems often interact, e.g.\ SHH, FGF10, and BMP together with their receptors regulate branching morphogenesis of the lung and several glands, while GDNF, FGF10, and WNT and their receptors regulate kidney branching morphogenesis, as recently reviewed \cite{Iber:OpenBiol:2013}. We were therefore interested how the interaction of several such Turing modules would affect the Turing space.

To that end, we carried out a systematic analysis of possible feedback interactions between two separate receptor-ligand-based Turing systems (for details see the Appendix II. E). The studied network architectures, systems of equations, and Turing spaces are shown in Fig.\ S1. Figure \ref{Fig1c}A summarizes the coupled Turing modules with the largest Turing spaces. Here, similar as for uncoupled modules (Fig.\ \ref{Fig1u}), the largest Turing space is observed when a negative feedback acts on the production rates (Fig.\ \ref{Fig1c}). We notice that coupling of the two Turing systems via a negative feedback on the constitutive receptor expression rates, $a$, results mainly in an increase in the parameter space of $a$ (Fig.\ \ref{Fig1c}B, blue), while coupling the two Turing systems via a negative feedback on the constitutive ligand production rate $b$ results mainly in an increase in the parameter space for $b$ (Fig.\ \ref{Fig1c}B, red). The asymmetrically coupled modules with one feedback on $a$ and one on $b$ have a very large (possibly infinitely large) parameter space (Fig.\ S1 panels C6, C8 and C10). However, the parameter range is very narrow and extends towards infinity only along the $b$-axis, while it is bounded above on the $a$-axis.  A massive increase in the size of the Turing space is observed when the two Turing modules are coupled by four negative feedbacks, such that all constitutive receptor and ligand expression rates are regulated by negative feedbacks (Fig.\ \ref{Fig1c}B, green and Fig. S1 panel C11). In this case, the parameter space dramatically increases in both directions as $p$ is lowered, such that already for $p=0.1$, the parameter ranges of both $a$ and $b$ expand by more than four orders of magnitude compared to a single receptor-ligand based Turing model, and further increase as $p$ is lowered (Fig.\ \ref{Fig1c}B, green and Fig. S1 panel C11).

\begin{figure}
\begin{center}
\includegraphics[width=\columnwidth]{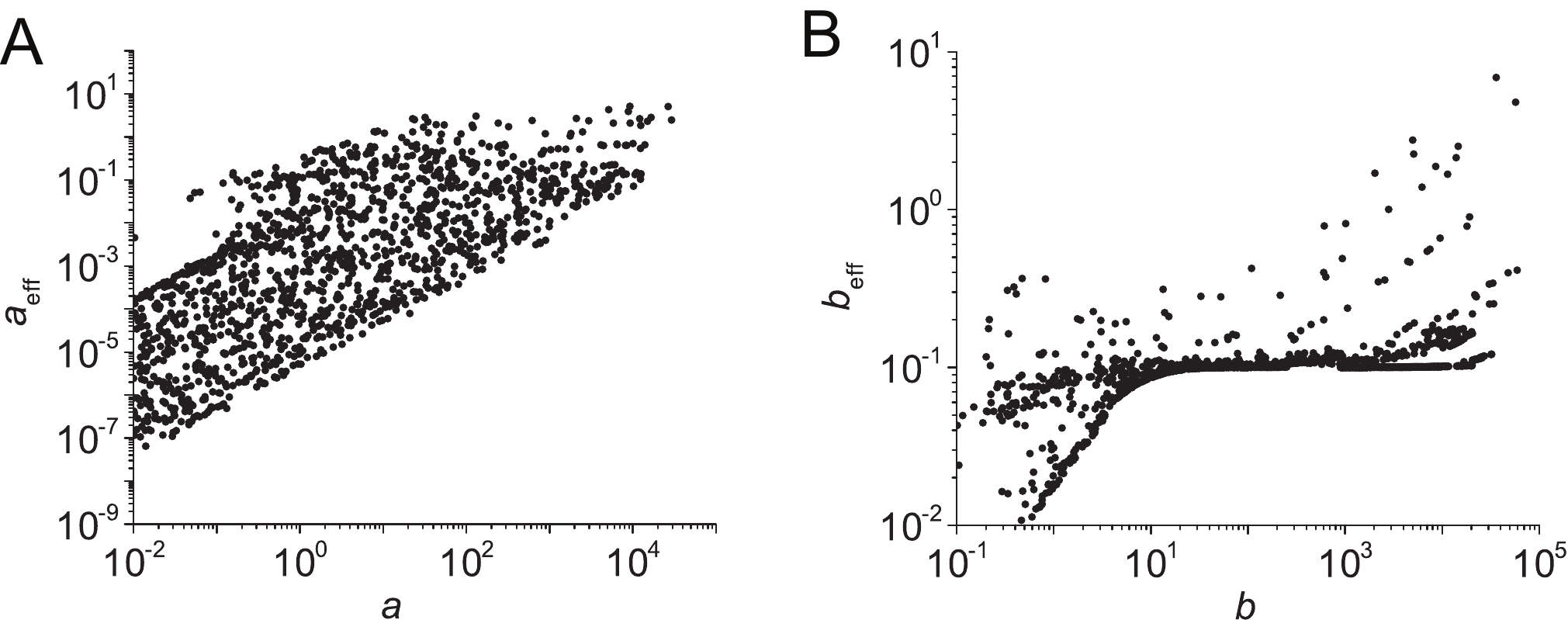}
\end{center}
\caption{
{\bf Negative Feedbacks enlarge the Turing space by limiting the effective production rates.}  The plot of the (A) effective receptor production rate $a_\mathrm{eff}=\frac{a}{\max (R^2 L) / p+1}$ versus the receptor production rate $a$, and (B)  the plot of the effective ligand production rate $b_\mathrm{eff}=\frac{b}{\max(R^2 L) / p+1}$ versus $b$ show that, as a result of the negative feedbacks, the effective production rates remain in a narrow range, even as $a$ and $b$ are greatly changed.  The calculation was carried out for the symmetrically coupled Turing system, shown in green in Figure \ref{Fig1c}B. }
\label{Fig1r}
\end{figure}

\subsection{Negative Feedbacks enlarge the Turing space by limiting the effective production rates}

We wondered why negative feedbacks would enlarge the Turing space. To this end, we plotted the effective production rates $a_\mathrm{eff}=\frac{a}{\max (R^2 L) / p+1}$ and $b_\mathrm{eff}=\frac{b}{\max(R^2 L) / p+1}$  for the coupled Turing systems with the largest Turing space (Fig.\ \ref{Fig1c}B, green) versus $a$ and $b$, respectively (Fig.\ \ref{Fig1r}). We find that the effective production rates are much smaller than what the parameter values $a$ and $b$ would suggest, and almost lie within the standard small Turing space. Thus, the negative feedback effectively corrects the receptor and ligand production rates,  and thereby enables the Turing mechanism to tolerate a much wider range of production rates.

\subsection{The restriction of receptors to single cells enlarges the Turing space}

\begin{figure}
\begin{center}
\includegraphics[width=\columnwidth]{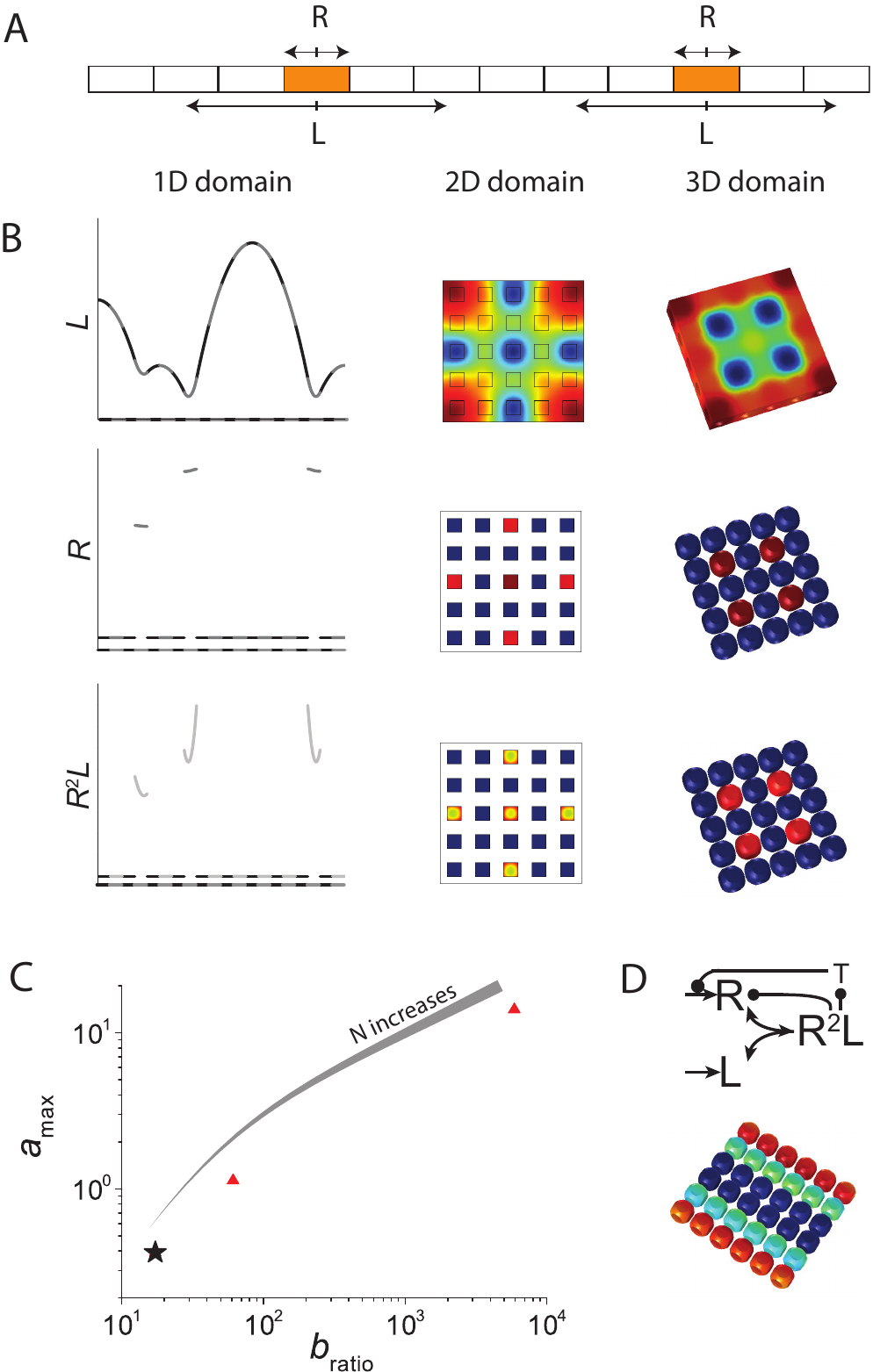}
\end{center}
\caption{
{\bf The restriction of receptors to single cells enlarges the Turing space.} (Color online) \small{ (A) Cartoon of the computational domain: diffusion of receptors is restricted to single cells, while ligand can diffuse over the entire computational domain. (B) Solution of the receptor-ligand model on a 1D, 2D and 3D (left to right) cellularized computational domain. The ligand (upper row), receptors (middle row), and ligand-receptor complexes (bottom row) pattern the domain. We provide the concentration levels (in arbitrary units) on the vertical axis for the 1D domain (left column), and intensities as colour code (blue- low; red - high) on the 2D and 3D domains. To distinguish cell boundaries on the 1D domain we alternate black and grey lines. (C) The size of the Turing space increases as the domain of fixed size is split into more cells, $N$. Triangles show the results for $N = 10$ and $N = 100$ cells. The black star reports the Turing space for the standard model, $N=1$.  (D) Patterns of receptor-ligand complexes that extend over several cells can be obtained with a diffusive component, $T$, that is produced in response to the formation of receptor-ligand complexes, and that enhances the abundance of receptors on neighbouring cells. The grey arrow indicates the direction, in which the feedback threshold, $p$, decreases.}
}
\label{Fig2}
\end{figure}

\begin{figure}
\begin{center}
\includegraphics[width=\columnwidth]{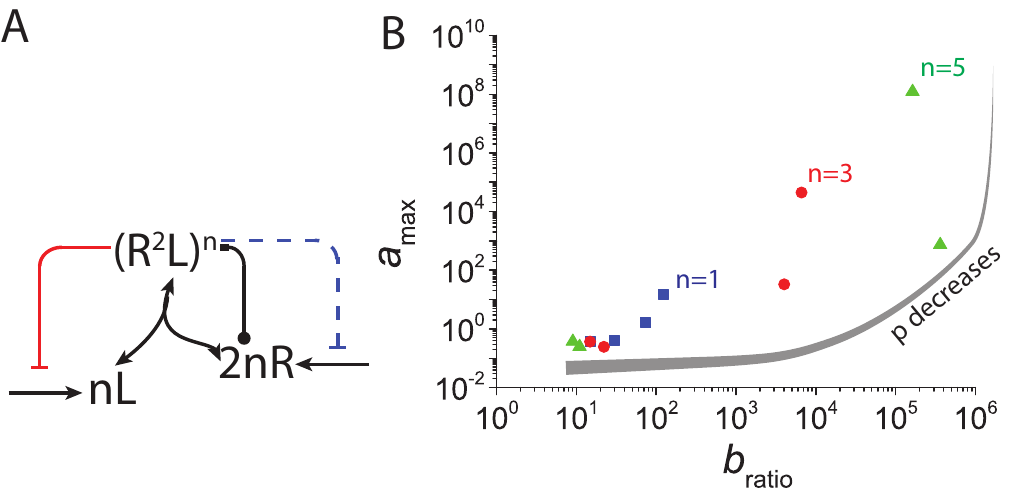}
\end{center}
\caption{
{\bf Receptor clustering enlarges the Turing space.}  (Color online)  (A) The simulated network architecture. Clusters of $2 n$ receptors $R$ interact with $n$ dimeric ligands $L$ to form a receptor-ligand complex $(R^2L)^n$ (black arrows, $\leftrightarrow$).  The receptor-ligand complex upregulates the presence of receptor (black interaction, $-\bullet$). In addition to these core interactions that can result in a Turing mechanism, we considered negative feedbacks on the ligand production (red, solid arrow, $\dashv$) and / or the receptor production (blue, dashed arrow, $\dashv$). (B) Higher cooperativity, $n>1$, as may result from larger receptor-ligand clusters further increases the size of the Turing space. The $n$-dependent increase was calculated for $p$ = 0.01, 0.1, 1, 10, 100 for case U5 in Fig. S1. The grey arrow indicates the direction, in which the feedback threshold, $p$, decreases. }
\label{Fig6}
\end{figure}

\begin{figure*}
\begin{center}
\includegraphics[width=0.8\textwidth]{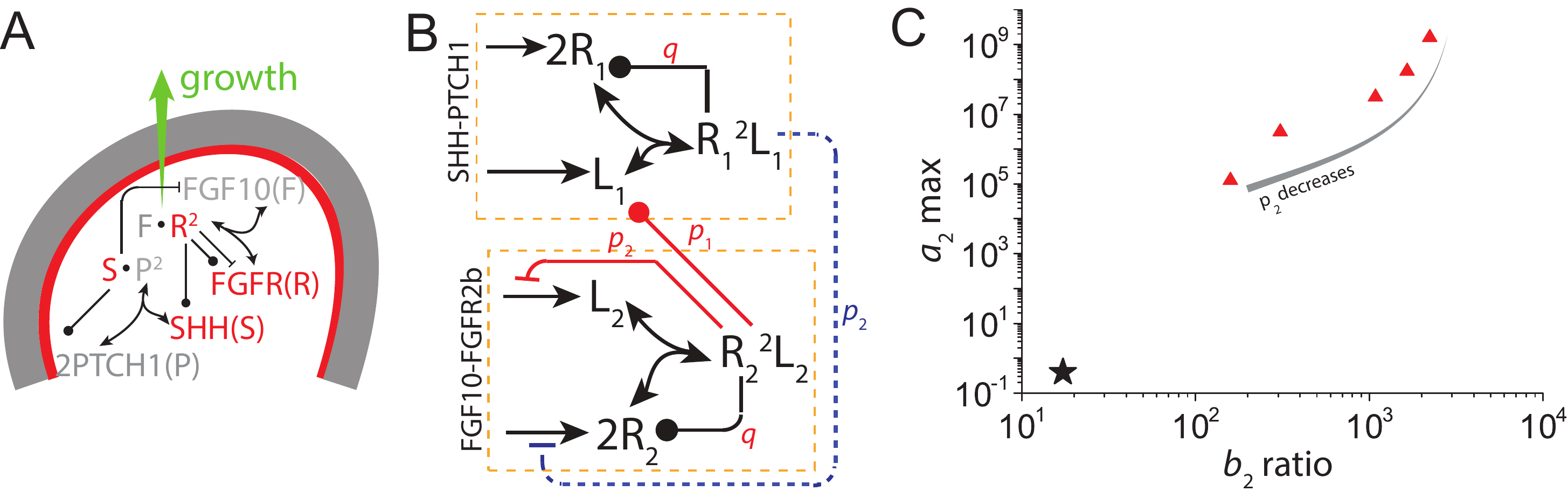}
\end{center}
\caption{
{\bf Substantially enlarged Turing spaces for physiological networks.}  (Color online)  (A) The SHH-FGF10 network in the control of lung branching morphogenesis. For details see text. (B) Schematic representation of the regulatory network for lung branching morphogenesis in panel A.  (C) The Turing space of such a physiological model is huge, and further increases as the feedback threshold, $p$, is lowered. The red triangles represent the Turing spaces for $p_1 = q = 0.1$ (positive feedback on ligand and receptor, respectively) and $p_2$ = 0.01, 0.1, 1, 10, 100 (negative feedback); the black star represents the size of the Turing space of the standard network in Fig. 2A (black part). The grey arrow indicates the direction, in which the feedback threshold, $p$, decreases. $\leftrightarrow$ indicates binding interactions, $\dashv$ indicates inhibitory interactions, and $-\bullet$ indicates up-regulating interactions.}
\label{Fig:physiol}
\end{figure*}

So far, we have treated receptors in the same way as the ligand, just with a  smaller diffusion coefficient. However, receptors are confined to single cells and thus cannot diffuse from one cell to the next. Moreover, they often cluster on the cell surface.  We therefore next studied Turing patterns on cellular domains where receptors are confined to single cells, while ligands can diffuse within the tissue (Fig. \ref{Fig2}A). The computational details of the implementation have previously been described \cite{Menshykau:ProceedingsOfComsolConference2012:2012}, and details  of the implementation are given in Appendix III. In brief, to restrict diffusion of receptors to a single cell in 1D and 2D models (Fig. \ref{Fig2} B, left and middle panel, respectively), we set no-flux boundary conditions for receptor at the pseudo-cell boundary, while ligand was free to diffuse in the entire domain. In the 3D model the cell surfaces were approximated as spheres (Fig. \ref{Fig2}B, right panel), and both ligands and receptors were produced on the spheres' surfaces. Diffusion of receptors was restricted to the surface of each sphere, while ligand was free to diffuse also in the intercellular space; the details of the implementation have been previously described \cite{Vollmer:CordConferenceProceedings:2013}. 

We observe the emergence of patterns on 1D, 2D and 3D cellularized domains (Fig. \ref{Fig2}B), and as a tissue domain of a given size is divided into more (and thus smaller) cells, to which the receptors are restricted, the Turing space increases (Fig. \ref{Fig2}C). Interestingly, however, cells with a high level of receptor-ligand complexes occur only as isolated spots (Fig. \ref{Fig2}B, red spots), while clusters of such active cells are not observed. To obtain clusters of active cells we have to include a second diffusively component, $T$, that is secreted by the active cells and that activates neighboring cells (Fig. \ref{Fig2}D).

\subsection{Receptor clustering enlarges the Turing space}

Receptors often cluster on cell membranes, either as pre-clusters or induced by multimeric ligand. Clustered receptor-ligand complexes may cooperate \cite{Bray:Nature:1998}, such that regulation is not mediated by a single ligand-receptor complex, but by the cluster. We then have $(R^2L/p)^n$ with $n>1$ in Eq.\ \eqref{eq:1}-\eqref{eq:2} instead of $R^2L/p$. As we increase $n$, we observe a further increase in the size of the Turing space (Fig.\ref{Fig6}A,B). In summary, both receptor clustering and the cellular restriction of receptors greatly increase the Turing space.

\subsection{Physiological Turing Models}
Physiological networks harbour many feedbacks and we wondered by how much the size of the Turing space would be increased in physiological settings. Here, we considered the network that controls branching morphogenesis in the lung (Fig. \ref{Fig:physiol}A); similar networks also operate in the prostate, salivary gland, and in the pancreas \cite{Iber:OpenBiol:2013}. Core to the control of lung branching morphogenesis are FGF10 and SHH as no branching is observed in the null mutants \cite{Abler:DevDyn:2009, Chiang:Nature:1996, Pepicelli:CurrBiol:1998, Bellusci:Development:1997b}, and expression of dominant negative \textit{Fgfr2} blocks lung branching, but not outgrowth \cite{Peters:EmboJ:1994}. 

FGF10 upregulates \textit{Shh} expression \cite{Abler:DevDyn:2009} and the expression of its own receptor, FGFR2b \cite{Cui:IntJMolSci:2013}, while SHH signaling downregulates \textit{Fgf10} expression\cite{Bellusci:Development:1997} and upregulates the expression of its own receptor \textit{Ptch1} \cite{Ingham:GenesDev:2001} (Fig.\ref{Fig:physiol}A). We have previously shown that the SHH-PTCH kinetics can be described by Eq.\ref{eq:1}-\ref{eq:2} \cite{Tanaka:PhysicalBiology:2013, Menshykau:PlosComputBiol:2012}; similar equations can also be derived for the FGF10-FGFR2b kinetics; see Appendix I for a general derivation of the ligand-receptor kinetics. The particular stoichiometry in Eq. \eqref{eq:1}-\eqref{eq:2} assumes the binding of one ligand dimer to two receptor monomers. In case of FGF10, monomeric binding of one FGF10 dimer to its trivalent FGFR2b receptor triggers dimerization of the FGF10-receptor complex \cite{Ibrahimi:MolecularAndCellularBiology:2005}; SHH is a multimer that may form higher order complexes with its receptor PTCH1 \cite{Zeng:Nature:2001}. We have previously shown that similar Turing patterns can be observed also with such very different stochiometries \cite{Menshykau:PlosComputBiol:2012}. For ease of comparison, we stick to the standard model (Eq. \ref{eq:1}-\ref{eq:2}) for the FGF10 and SHH modules, though we note that larger SHH/PTCH1 clusters would further increase the Turing space (Fig. \ref{Fig2}E,F). The two signalling factors interact in that FGF10 upregulates \textit{Shh} expression \cite{Abler:DevDyn:2009}, while SHH signaling downregulates \textit{Fgf10} expression\cite{Bellusci:Development:1997}. The equations for the coupled network (Fig.\ref{Fig:physiol}B) are thus given by

\begin{eqnarray}\label{Eq_lung_network}
\textup{PTCH1:} \qquad & &  \dot{R_1} = \Delta R_1 + f(R_1,L_1,R_2,L_2)  \nonumber \\ 
\textup{SHH:} \qquad & &  \dot{L_1} = d \Delta L_1 +g(R_1,L_1,R_2,L_2)   \nonumber \\
\textup{FGFR2b:} \qquad  & & \dot{R_2} = \Delta R_2 + \widetilde f(R_1,L_1,R_2,L_2) \nonumber \\
\textup{FGF10:} \qquad  & &  \dot{L_2} = d \Delta L_2 + \widetilde g(R_1,L_1,R_2,L_2) 
\end{eqnarray}

with

\begin{eqnarray}\label{Eq_lung_network_detailed}
 f(R_1,L_1,R_2,L_2) & =& a_1 - R_1  + qR_1^2L_1 \nonumber \\ 
g(R_1,L_1,R_2,L_2) &  =& b_1- R_1^2L_1 + p_1R_2^2L_2 \nonumber  \\
\widetilde f(R_1,L_1,R_2,L_2) & = &\frac{a_2}{1+\frac{R_2^2L_2}{p_2}} - R_2 + qR_2^2L_2 \nonumber \\
\widetilde g(R_1,L_1,R_2,L_2) &  = &\frac{b_2}{1+\frac{R_1^2L_1}{p_2}} - R_2^2L_2
\end{eqnarray}

Here $R_1$ represents the receptor PTCH1, $L_1$ the ligand SHH, $R_2$ the receptor FGFR2b, and $L_2$ the ligand FGF10. The SHH-PTCH1 complex, $R_1^2L_1$ upregulates the receptor PTCH1 \cite{Ingham:GenesDev:2001}, i.e.\ $+ qR_1^2L_1$ in the term $ f(R_1,L_1,R_2,L_2) $, and inhibits the production of FGF10 \cite{Bellusci:Development:1997}, i.e.\ $\frac{b_2}{1+\frac{R_1^2L_1}{p_2}}$ in the term $\widetilde g(R_1,L_1,R_2,L_2)$. The FGF-receptor complex, $R_2^2L_2$, upregulates the production of SHH, i.e. $p_1R_2^2L_2$ in the term $g(R_1,L_1,R_2,L_2) $ and both upregulates \cite{Cui:IntJMolSci:2013}, i.e.\ $qR_2^2L_2$ in term $\widetilde f(R_1,L_1,R_2,L_2) $, and downregulates \cite{Monzat:JournalOfBiologicalChemistry:1996}, i.e.\ $\frac{a}{1+\frac{R_2^2L_2}{p_2}}$ in term $\widetilde f(R_1,L_1,R_2,L_2) $, the FGF receptor FGFR2b. The $- R_1^2L_1$ and $- R_2^2L_2$ terms represent ligand removal by receptor binding; receptor removal by ligand binding is absorbed in the $ + qR_1^2L_1$ and $+ qR_2^2L_2 $ terms as signalling-dependent receptor upregulation dominates ligand-induced receptor removal. 

We find that the combination of these two modules (Fig. \ref{Fig:physiol}B) increases the range of the receptor production rate, $a$, by about $10^9$-fold as the threshold $p$ is lowered to 0.01, while the relative range of the ligand production rate, $b_2$, increases about 100-fold compared to the single receptor-ligand based Turing model (Fig.\ \ref{Fig:physiol}C). \\

\section{DISCUSSION} 

Turing mechanisms can reproduce a wide range of biological patterning phenomena. However, it has remained unclear how they may be implemented on the molecular scale and how they could evolve in spite of the small sizes of their Turing spaces. We propose that ligand-receptor interactions give rise to Turing patterns, and we show that negative feedbacks, the coupling of Turing modules, and the restriction of receptors to single cells can greatly increase the size of the Turing space (Fig. \ref{Fig1u}, \ref{Fig1c}, \ref{Fig2}) and thus increase the range of parameter values for which Turing patterns will emerge in biological systems. 

The conditions for ligand-receptor based Turing mechanisms, as summarised in the Introduction, are met by many different ligand-receptor pairs, and we have previously shown that receptor-ligand based Turing mechanisms can indeed well describe the patterning processes for a range of developmental systems \cite{Menshykau:PhysicalBiology:2013, Tanaka:PhysicalBiology:2013, Menshykau:PlosComputBiol:2012, Badugu:SciRep:2012,Iber:OpenBiol:2013}. Equally, negative feedbacks are prevalent in biological regulation and have previously been shown to enable robustness to noise \cite{Becskei:Nature:2000} and transient responsiveness \cite{Ma:Cell:2009}. We now propose that negative feedbacks enable robust patterning also for receptor-ligand based Turing mechanisms. Interestingly, also the restriction of receptors to single cells can further increase the size of the Turing space (Fig. \ref{Fig2}). This suggests a way of how Turing mechanisms may have first evolved. Cooperative interactions in receptor clusters and the introduction of feedbacks as well as the coupling of several Turing modules may then have further increased the size of the Turing space. 

It will be important to test our theoretical insights by synthetically constructing such ligand-receptor-based Turing mechanism, and by establishing the key parameter values (rates of production, decay, diffusion coefficients, endogenous concentrations etc.) in the living systems. The Turing space of ligand-receptor systems with additional negative feedbacks should be sufficiently large that synthetic biology approaches can now obtain Turing patterns in spite of the difficulties to accurately control kinetic rates in synthetic biology approaches. Given their robustness and flexibility, we propose that receptor-ligand based Turing mechanisms are the likely standard way how Turing mechanisms are implemented in Nature.\\


\section{Materials \& Methods}
The Turing space was defined based on a Linear Stability Analysis  \cite{Murray:MathematicalBiology3RdEditionIn2Volumes:2003} as described in Appendix II, using MATLAB. The implementation of cell-based models is described in Appendix III. The partial differential equations were solved in COMSOL Multiphysics 4.3 as described previously \cite{Vollmer:CordConferenceProceedings:2013, Germann:ProceedingsOfComsolConference2011:2011, Menshykau:ProceedingsOfComsolConference2012:2012}. Tests of the numerical methods are provided in Appendix IV.  \\

\section{Acknowledgments}
The authors thank Patrick Fried and Jannik Vollmer for discussions.

\section*{Appendix I: Derivation of the Equations for the Receptor-Ligand Signaling Model} 

As previously  derived \cite{Badugu:2012ho,Menshykau:2012kg,Tanaka:2013wt}, the dynamics of receptors, $R$, ligands, $L$, and the ligand-receptor complex, $C$ (Fig. 1C), can be described by the following set of equations:
\begin{widetext}
\begin{eqnarray} \label{eqL}
\dot{\![\mathrm{L}]}  =  \underbrace{\overline{D}_\mathrm{L} \overline{\Delta} [\mathrm{L}]}_{\text{\ diffusion}} + \underbrace{ \overline{\rho}_{\mathrm{S}}}_{\text{\ production}} \underbrace{-\overline{\delta}_{\mathrm{L}}[\mathrm{L}]}_{\text{\ degradation}} \underbrace{-n\, \overline{\mathrm{k}}_\mathrm{on}[\mathrm{R}]^m[\mathrm{L}]^n +n \overline{k}_\mathrm{off}[\mathrm{C}]}_{\text{\ complex formation}}  
\end{eqnarray}

\begin{eqnarray} \label{eqR}
\dot{\![\mathrm{R}]} &=& \underbrace{\overline{D}_\mathrm{R} \overline{\Delta} [\mathrm{R}]}_{\text{\ diffusion}} +\underbrace{ \overline{\rho}_{\mathrm{R}}+\mu([\mathrm{C}])}_{\text{\ production}} \underbrace{-\overline{\delta}_{\mathrm{P}}[\mathrm{P}]}_{\text{\ degradation}} \underbrace{-m\, \mathrm{k}_\mathrm{on}[\mathrm{R}]^m[\mathrm{L}]^n +m \overline{k}_\mathrm{off}[\mathrm{C}]}_{\text{\ complex formation}}  
\end{eqnarray}

\begin{eqnarray} \label{eqC}
\dot{\![\mathrm{C}]} &=& \underbrace{\overline{D}_\mathrm{C} \overline{\Delta} [\mathrm{C}]}_{\text{\ diffusion}}+\underbrace{\, \mathrm{k}_\mathrm{on}[\mathrm{R}]^m[\mathrm{L}]^n - \overline{k}_\mathrm{off}[\mathrm{C}]}_{\text{\ complex formation}}  \underbrace{-\overline{\delta}_{\mathrm{C}}[\mathrm{C}]}_{\text{\ degradation}}. 
\end{eqnarray}
\end{widetext}
Here $[X]$ denotes the concentration of component $X$, $\overline{D}_X$ denotes the diffusion coefficient, $\overline{\rho}_X$ the production rate constant, and $\overline{\delta}_X$ the first order degradation rate constant of component $X$. $\mu([C])$ specifies a function that describes the ligand-receptor dependent up regulation of receptor production. $k_\mathrm{on}$ denotes the rate constant for the formation, and $k_\mathrm{off}$ the rate constant for the dissociation of the ligand-receptor complex. $m$ and $n$ specify the number of receptors and ligands that bind in the ligand-receptor complex.\\

Assuming that the dynamics of the complex are fast compared to those of the other components, we can  introduce a quasi-steady state approximation
\begin{eqnarray} \label{eqC}
0 &=& \underbrace{\, \mathrm{k}_\mathrm{on}[\mathrm{R}]^m[\mathrm{L}]^n - \overline{k}_\mathrm{off}[\mathrm{C}]}_{\text{\ complex formation}}  \underbrace{-\overline{\delta}_{\mathrm{C}}[\mathrm{C}]}_{\text{\ degradation}},
\end{eqnarray}
and thus arrive at the quasi-steady state concentration of complex [C]$_\mathrm{SS}$
\begin{eqnarray} \label{eqC_qstst}
[\mathrm{C}] _\mathrm{SS}= \frac{\overline{k}_\mathrm{on}}{\overline{k}_\mathrm{off} + \overline{\delta}_{\mathrm{C}}} [\mathrm{R}]^m [\mathrm{L}]^n = \overline{\Gamma} [\mathrm{R}]^m [\mathrm{L}]^n,
\end{eqnarray}
where $\overline{\Gamma}= \frac{\overline{k}_\mathrm{on}}{\overline{k}_\mathrm{off} + \overline{\delta}_{\mathrm{C}}}$. The concentration of bound receptor, [C],  is thus proportional to $[\mathrm{R}]^m\mathrm{[S]^n}$. Furthermore, assuming that the rate of receptor upregulation in response to receptor-ligand signaling $\mu([\mathrm{C}]) =  \overline{v} [\mathrm{C}]= \overline{v} \overline{\Gamma}[\mathrm{R}]^m[\mathrm{L}]^n $ depends linearly on the ligand-receptor complex concentration, $[C]$, we obtain the following set of PDEs:
\begin{eqnarray} \label{eq:Sf}
\dot{\![\mathrm{L}]} &=& \overline{D}_\mathrm{L}\overline{\Delta}  [\mathrm{L}] + \overline{\rho}_{\mathrm{L}}  -  n\overline{\delta}_{\mathrm{C}}\overline{\Gamma} [\mathrm{R}]^m[\mathrm{L}]^n   -\overline{\delta}_{\mathrm{L}}[\mathrm{L}] 
\end{eqnarray}
\begin{eqnarray}\label{eq:Pf}
\hspace*{-0.2cm} \dot{\![\mathrm{R}]} &=& \overline{D}_\mathrm{R}\overline{\Delta} [\mathrm{P}] + \overline{\rho}_\mathrm{R}+ (\overline{v} -m \overline{\delta}_\mathrm{C})\overline{\Gamma}[\mathrm{R}]^m[\mathrm{L}]^n - \overline{\delta}_\mathrm{R}[\mathrm{ R}]
\end{eqnarray}
We note that the linear response of the receptor production rate to receptor-ligand signaling helps to increase the size of the Turing space. Based on the results in Fig.\ S1, case U6, we expect that a saturation of the response for higher ligand-receptor concentrations, as could be described by a Hill function of the form $\mu([\mathrm{C}]) =H(\mu([\mathrm{C}], K)= H(\mathrm{\overline{\Gamma} [\mathrm{R}]^n[\mathrm{L}]^m},K)$, would cause a shrinking of the Turing space.

Equations \eqref{eq:Sf}-\eqref{eq:Pf} converge to the classical Schnakenberg equations for the following conditions: 
\begin{itemize}
\item Receptor-independent degradation of ligand is much less efficient than receptor-dependent ligand degradation, as is generally the case, i.e.\ $\overline{\delta}_{\mathrm{L}}[\mathrm{L}]  \ll  n\overline{\delta}_{\mathrm{C}}\overline{\Gamma}\mathrm{R}]^m[\mathrm{L}]^n$.
\item The stochiometry of the ligand-receptor interaction yields $m$=2, $n$=1; we note that other stochiometries also yield Turing patters  \cite{Menshykau:2012kg}. 
\item $\overline{v}  = (m+ n) \overline{\delta}_\mathrm{C}$. 
\end{itemize}


\subsection{Derivation of the Non-Dimensional Set of Equations for the Receptor-Ligand based Turing Mechanism}

In the following, we will adopt the standard notation that is used to describe Turing models, and we write $U$ for the receptor concentration and $V$ for the ligand concentration; $U^mV^n$ represents the quasi-steady state concentration of the receptor-ligand complex. We have previously shown that a wide range of stochiometries can yield Turing patters \cite{Menshykau:2012kg}. Using $m=2$, $n=1$, i.e.\ one ligand dimer $V$ binds to two monomeric receptors $U$, equations \eqref{eq:Sf}-\eqref{eq:Pf} can be written as 

\begin{eqnarray}
\frac{\partial U}{\partial\tau} & =&  D_\mathrm{U}\Delta U +k_1-k_2 U +(k_5-2 k_3) U^2V \label{Schnakenberg_U_dim} \\[2mm]
\frac{\partial V}{\partial\tau} & = &  D_\mathrm{V}\Delta V +k_4 -k_6 V - k_3 U^2V, \label{Schnakenberg_V_dim}
\end{eqnarray}
where $U=U(\tau,{\bf X})$ and $V=V(\tau,{\bf X})$ are the unknown functions depending on the time variable $\tau$ and space variable ${\bf X}$. The coefficient $k_1$ then represents the constitutive receptor production rate, while $k_4$ represents the constitutive ligand production rate. The term $-k_2 U$ reflects the ligand-independent receptor turn-over rate while $-k_6 V$ reflects the receptor-independent ligand turn-over rate. $-k_3 U^2V$ reflects the turn-over of the receptor-ligand complex, which leads to the removal of one ligand dimer, $V$, and two receptor monomers, $U$. Most ligand is typically removed by this receptor-dependent process, and we can therefore make the simplifying approximation $k_6=0$. Finally, $+k_5 U^2V$ reflects the signalling-dependent increase in receptor emergence (which can happen by a wide range of mechanisms); we will set $k_5=3 k_3$ in the following to recover the classical Schnakenberg equations.  Equations \eqref{Schnakenberg_U_dim}-\eqref{Schnakenberg_V_dim} then read
\begin{eqnarray}
\frac{\partial U}{\partial\tau}& = &D_\mathrm{U}\Delta U +k_1-k_2 U + k_3 U^2V\\[2mm]
\frac{\partial V}{\partial\tau}& = & D_\mathrm{V}\Delta V +k_4 - k_3 U^2V.
\end{eqnarray}
These equations can be rewritten in dimensionless form as 
\begin{equation}\label{eq:Schnakenberg}
\begin{split}
\frac{\partial u}{\partial t}&=\Delta u +\gamma(a-u+u^2v)\\[2mm]
\frac{\partial v}{\partial t}&=d\Delta v +\gamma(b -u^2v),
\end{split}
\end{equation}
where
\begin{gather*}
u = U\left(\frac{k_3}{k_2}\right)^{1/2},\ v = V\left(\frac{k_3}{k_2}\right)^{1/2},\ t = \frac{D_\mathrm{U}\,\tau}{L^2},\ {\bf x} = \frac{{\bf X}}{L},\\ d=\frac{D_\mathrm{V}}{D_\mathrm{U}},\, a = \frac{k_1}{k_2} \left(\frac{k_3}{k_2}\right)^{1/2},\, b = \frac{k_4}{k_2}\left(\frac{k_3}{k_2}\right)^{1/2},\, \gamma = \frac{L^2 k_2}{D_\mathrm{U}}.
\end{gather*}

The function $u$ then represents the receptor, $v$ represents the ligand, and $u^2v$ represents the quasi-steady state concentration of the receptor-ligand complex. As before, one ligand dimer $v$ binds to two monomeric receptors $u$. We have previously shown that also other combinations $u^m v^n$ result in Turing patterns \cite{Menshykau:2012kg}. The constant $\gamma a$ then represents the constitutive receptor production rate, while $\gamma b$ represents the constitutive ligand production rate. The term $-\gamma u$ reflects the ligand-independent receptor turn-over rate, while $-\gamma u^2v$ reflects the receptor-dependent ligand removal rate. Finally, $+\gamma u^2v$ represents the net result of ligand-dependent receptor turn-over and the signalling-dependent increase in receptor emergence, where the latter dominates, thus the positive term.

\section*{Appendix II: Determination of Turing Spaces}
\subsection{The Turing Mechanism}

\label{sec:Tpf}

In this section we summarize briefly the criteria for the emergence of Turing pattern  for reaction-diffusion systems with two species. We consider systems of the form
\begin{equation}
\label{eq:generalReDiffEq}
\begin{split}
\frac{\partial U}{\partial \tau} & = F(U, V) + D_U\Delta U \\[2mm]
\frac{\partial V}{\partial \tau} & = G(U, V) + D_V\Delta V
\end{split}
\end{equation}
defined on $(0,\infty)\times \Omega$ (with a given spatial domain $\Omega\subset\mathbb{R}^n$) subject to boundary and initial conditions, where the space and time-dependent functions $U$ and $V$ represent concentrations and the reaction kinetic terms $F$ and $G$ are generally nonlinear functions. After suitable changes of variables and nondimensionalization Eq.\ \eqref{eq:generalReDiffEq} can be transformed into the dimensionless system
\begin{equation}
\label{eq:nondimReDiffEq}
\begin{split}
u_t & = \gamma f(u, v) + \Delta u \\
v_t & = \gamma g(u, v) + d\Delta v,\\
\end{split}
\end{equation}
where $t$ is the rescaled time variable, $d$ denotes (or is proportional to) the quotient of the diffusion coefficients $D_U$ and $D_V$ and $\gamma = \textit{const} \cdot L^2$, where $L$ is a typical length scale of the domain. To ensure the uniqueness of the solution we endow system \eqref{eq:nondimReDiffEq} with initial and boundary conditions. We will use homogeneous Neumann boundary condition of the form
\begin{equation*}
\begin{split}
({\bf n}\cdot\nabla)\begin{pmatrix}u\\ v\end{pmatrix} = 0 \quad \mbox{on } [0,\infty)\times\partial\Omega \\
\qquad u(0,{\bf x}) = u_0({\bf x}),\ v(0,{\bf x}) = v_0({\bf x}),
\end{split}
\end{equation*}
because they are easy to handle and have a biological interpretation (impermeable boundary). We note, however, that other boundary conditions would not greatly alter the following analysis. A Turing instability appears when a reaction-diffusion system has a stable steady state in the absence of diffusion, which loses its stability in the presence of diffusion such that spatial patterns emerge.

\subsection{Linear stability in the absence of diffusion}

Let $u_0$ and $v_0$ denote the steady state of the diffusion-free system of ordinary differential equations (ODEs)
\begin{equation}
\label{eq:generalODE}
u_t = \gamma f(u,v),\qquad v_t = \gamma g(u,v),
\end{equation}
and linearize the system about $(u_0,v_0)$ by introducing the translated function ${\bf w} = (w_1, w_2)^T$ with $w_1 = u-u_0,\ w_2 = v-v_0$. Then the linearized system becomes
\[
{\bf w}_t = \gamma J{\bf w},
\]
where
\[
J = \begin{pmatrix} f_u & f_v\\ g_u & g_v\end{pmatrix}\Bigg|_{(u_0,v_0)} = \begin{pmatrix} f_u(u_0,v_0) & f_v(u_0,v_0)\\ g_u(u_0,v_0) & g_v(u_0,v_0)\end{pmatrix}
\]
is the Jacobian evaluated at the point $(u_0,v_0)$. From now on, we write the partial derivatives evaluated at the steady state without their arguments for brevity. The steady state of the linearized system is stable, i.e.\ the steady state of system \eqref{eq:generalODE} is linearly stable if $\Re \lambda(J) < 0$ for all eigenvalues of $J$ (see any textbook on ODEs), which for a 2-component system is ensured by the conditions
\begin{equation}
\label{eq:linstab}
\textit{tr} J = f_u + g_v < 0,\qquad \textit{det}(J) = f_ug_v - f_vg_u > 0.
\end{equation}

\subsection{Diffusion-driven instability}
\label{subsec:diffdrivinstab}

Now let us add diffusion to our system of ODEs and consider the reaction-diffusion system linearized about the steady state ${\bf w} = (0,0)^T$, which has the form
\begin{equation}
\label{eq:linwdiff}
{\bf w}_t = \gamma J{\bf w} + D\Delta {\bf w},
\end{equation}
where $D = \textit{diag}(1,d)$ is a diagonal matrix containing the diffusion coefficients of the nondimensionalized system \eqref{eq:nondimReDiffEq}. We look for a solution of the form
\begin{equation}
\label{eq:w}
{\bf w}(t, {\bf x}) = \sum_k {\bf C}_k \textup{e}^{\lambda_k t}{\bf W}_k({\bf x}),
\end{equation}
where the exponents $\lambda_k$ determine the temporal growth of the solution and the time-in\-de\-pen\-dent functions ${\bf W}_k$ are the solutions of the elliptic eigenvalue problem
\[
\Delta {\bf W}_k + k^2{\bf W}_k = 0,\qquad ({\bf n}\cdot\nabla){\bf W}_k = 0.
\]
For instance, in one dimension on the interval $[0,L]$ the eigenvalues are $k = n\pi/L\ (n=0,1,2,\ldots)$, also called wavenumbers, and the eigenfunctions are $W(x) = \cos(n\pi x/L) = \cos(kx)$. The constants ${\bf C}_k = (C^{(1)}_k,C^{(2)}_k)^T$ are the Fourier-coefficients of the initial conditions.

Inserting equation \eqref{eq:w} into equation \eqref{eq:linwdiff} and using the fact that the set of eigenfunctions of the Laplace operator $\{{\bf W}_k\}$ forms a complete orthonormal system, we obtain as linearized system
\begin{equation}
\label{eq:linwdiff}
{\bf w}_t = \gamma J{\bf w} + D k^2 \bf{w}
\end{equation}
for each wavenumber $k$. Writing
\[
\det(\lambda I - \gamma J +  k^2 D) = 0,
\]
where $I=I_2$ is the $2$-by-$2$ identity matrix, we obtain the eigenvalues $\lambda = \lambda_k$ of the matrix $M=\gamma J - k^2 D$. Expanding the above determinant, we obtain that $\lambda_k$ is the root of the second order polynomial equation
\begin{widetext}
\[
\lambda^2 + \lambda\big(k^2(1+d) - \gamma(f_u+g_v)\big) + dk^4 - \gamma(df_u + g_v)k^2 + \gamma^2(f_ug_v-f_vg_u)=0.
\]
\end{widetext}
Since we look for unstable solutions, we require that $\Re \lambda_k > 0$ for some $k\ne 0$. This means that either the coefficient of $\lambda$ and/or the constant term must be negative. Since the steady state is required to be linearly stable in the absence of diffusion (which corresponds to the case $k=0$), we must have $k^2(1+d) - \gamma(f_u+g_v) > 0$. Hence, to obtain a $\lambda$ with positive real part in the presence of diffusion we require
\[
h(k^2):= dk^4 - \gamma(df_u + g_v)k^2 + \gamma^2(f_ug_v-f_vg_u) < 0
\]
for some nonzero wavenumber $k$. Since we require $f_ug_v-f_vg_u>0$ for linear stability in the absence of diffusion ($k=0$) \eqref{eq:linstab}, it follows that $df_u + g_v > 0$ must hold. This condition is not sufficient to ensure the negativity of the function $h$; an elementary calculation shows that the minimum of $h$ is attained at the point
\[
k^2_{\textup{m}} = \gamma\frac{df_u+g_v}{2d},
\]
and the minimum value of $h$ is
\[
h_{\textup{min}} = h(k_{\textup{m}}^2) = \gamma^2\left[(f_ug_v-f_vg_u) - \frac{(df_u+g_v)^2}{4d}\right],
\]
which is negative if the expression in the bracket is negative.

\bigskip

In summary, the well-known conditions (see \cite[Sec.\ 2.3]{MurrayBook}) for which a reaction-diffusion system with two species exhibits a Turing instability are as follows:
\begin{equation}
\label{eq:Turingcond}
\begin{split}
f_u + g_v < 0,\quad f_ug_v-f_vg_u > 0,\\
df_u + g_v >0,\quad (df_u+g_v)^2 - 4d(f_u+g_v-f_vg_u) > 0,
\end{split}
\end{equation}
where all partial derivatives are evaluated at the steady state $(u_0,v_0)$. We note that it is possible that these conditions are satisfied, but that no pattern emerges. This is the case when $h$ is not negative for any $k$ within the discrete set of wavenumbers, and only takes a negative value in between two of these discrete wavenumbers. The distance between wavenumbers shrinks as $\gamma$ is increased, and in the limit of infinite $\gamma$ the spectrum of $k$ is continuous. Since $\gamma$ is related to the size of the spatial domain, it follows that on small domains pattern formation may not happen, while on a sufficiently increased domain patterns may be observed.

\subsection{Turing instability in interacting systems}
\label{sec:interacting}

We now consider two identical reaction-diffusion systems, which we couple with each other in several ways. When the couplings are of the same type (i.e. when the first 2-component Turing system based on $u$ and $v$ is coupled with the second Turing system that is based on $\widetilde u$ and $\widetilde v$ via the same functions $f$ and $g$), then we can derive exact conditions for the Turing instability, as an extension of the classical results that were presented in Section \ref{sec:Tpf} (see \cite[Sec.\ 2.3]{MurrayBook} for more details). For this let us consider systems of the form
\begin{equation}
\label{eq:coupledsys}
\begin{split}
 u_t & =  \gamma f(u,v,\widetilde u, \widetilde v)  + \Delta u \\ 
 v_t & =  \gamma g(u,v,\widetilde u, \widetilde v)  + d\Delta v \\ 
 \widetilde u_t & =  \gamma f(\widetilde u, \widetilde v, u,v) + \Delta \widetilde u \\ 
 \widetilde v_t & =  \gamma g(\widetilde u, \widetilde v, u, v)  + d\Delta \widetilde v, 
\end{split} 
\end{equation}
where the functions $f$ and $g$ describe the chemical reactions, $\gamma>0$ is a constant depending on the size of the domain and $d>0$ is a diffusion parameter. Let $(u_0,v_0,\widetilde u_0,\widetilde v_0)$ denote the steady state (assuming that there is only one, or at least they are isolated) of this system in the absence of diffusion (note that due to the symmetry $u_0 = \widetilde u_0$ and $v_0 = \widetilde v_0$) and -- just as in the uncoupled case \eqref{eq:generalODE} -- linearize the system about the steady state. The linearized system has the form
\[
{\bf w}_t = \gamma J{\bf w},
\] 
where
\[
J = \begin{pmatrix}f_u & f_v & f_{\widetilde u} & f_{\widetilde v} \\ g_u & g_v & g_{\widetilde u} & g_{\widetilde v} \\  f_{\widetilde u} & f_{\widetilde v} & f_u & f_v  \\ g_{\widetilde u} & g_{\widetilde v} & g_u & g_v \end{pmatrix}
\]
is the Jacobian matrix. Note the symmetry in $J$ that arises for this particular coupling. In this linearized system the steady state is stable if $\Re \lambda(J) < 0$ for all eigenvalues of $J$. The eigenvalues are the roots of the characteristic polynomial $k_J(\lambda)$ of $J$, which is now a fourth order polynomial for the coupled system. Due to the very special form of the coupling and the resulting symmetries in $J$, the polynomial $k_J$ can be factorized as
\begin{widetext}
\begin{equation}
\label{eq:kJ}
\begin{split}
k_J(\lambda)  = & \left[\lambda^2 + \lambda(-f_u-g_v-f_{\widetilde u}-g_{\widetilde v}) + f_ug_v-f_vg_u + f_ug_{\widetilde v} - f_{\widetilde v}g_u + f_{\widetilde u}g_v - f_vg_{\widetilde u} + f_{\widetilde u}g_{\widetilde v} - f_{\widetilde v}g_{\widetilde u}\right] \times \\ 
 & \left[\lambda^2 + \lambda(-f_u-g_v+f_{\widetilde u}+g_{\widetilde v}) + f_ug_v-f_vg_u - f_ug_{\widetilde v} + f_{\widetilde v}g_u - f_{\widetilde u}g_v + f_vg_{\widetilde u} + f_{\widetilde u}g_{\widetilde v} - f_{\widetilde v}g_{\widetilde u} \right].
\end{split}
\end{equation}
\end{widetext}
Hence $\Re \lambda(J) <0$ holds for all the four eigenvalues of $J$, that is the steady state of \eqref{eq:coupledsys} is linearly stable if both of the factors in \eqref{eq:kJ} have only roots with negative real part, i.e.\
\begin{widetext}
\begin{equation}
\label{eq:coupledlinstab}
f_u + g_v < \pm (f_{\widetilde u} + g_{\widetilde v}),\qquad f_ug_v - f_vg_u + f_{\widetilde u}g_{\widetilde v} - f_{\widetilde v}g_{\widetilde u} > \pm(f_ug_{\widetilde v} - f_{\widetilde v}g_u + f_{\widetilde u}g_v - f_vg_{\widetilde u}).
\end{equation}
\end{widetext}
Following the course of the uncoupled case, by adding diffusion, we again arrive at Eq.\ \eqref{eq:linwdiff}, now with the diffusion matrix $D = \textit{diag}(1,d,1,d)$. As before, we look for a solution of the form of Eq.\ \eqref{eq:w}. To this end, we determine the eigenvalues $\lambda = \lambda_k$ for $M=\gamma J -k^2 D$. The characteristic polynomial of this matrix -- given the special forms of $J$ and $D$ -- can be factorized as the product of two second order polynomials:
\begin{widetext}
\begin{equation}
\label{eq:kM}
\begin{split}
k_M(\lambda)  = & \left[\lambda^2 + \lambda\left(k^2(1+d) - \gamma(f_u+g_v+f_{\widetilde u}+g_{\widetilde v})\right) + dk^4 - \gamma k^2(df_u + g_v + df_{\widetilde u}+ g_{\widetilde v}) + \right. \\ & \qquad \quad\left. \gamma^2(f_ug_v-f_vg_u + f_ug_{\widetilde v} - f_{\widetilde v}g_u + f_{\widetilde u}g_v - f_vg_{\widetilde u} + f_{\widetilde u}g_{\widetilde v} - f_{\widetilde v}g_{\widetilde u})\right] \times \\ 
 & \left[\lambda^2 + \lambda\left(k^2(1+d) - \gamma(f_u+g_v - f_{\widetilde u}-g_{\widetilde v})\right) + dk^4 - \gamma k^2(df_u + g_v - df_{\widetilde u} - g_{\widetilde v}) +\right. \\ & \qquad \quad\left.  \gamma^2(f_ug_v-f_vg_u - f_ug_{\widetilde v} + f_{\widetilde v}g_u - f_{\widetilde u}g_v + f_vg_{\widetilde u} + f_{\widetilde u}g_{\widetilde v} - f_{\widetilde v}g_{\widetilde u})\right]. 
\end{split}
\end{equation}
\end{widetext}
To obtain a Turing instability, at least one of the roots of $k_M$ has to have positive real part for some $k\ne 0$, i.e.\ one of the factors of $k_M$ must have a root with $\Re \lambda(M)>0$. The first factor of \eqref{eq:kM} has a root with positive real part if the coefficient of $\lambda$ is negative, or the constant term is negative. But since the steady state is stable in the absence of diffusion (linear stability conditions \eqref{eq:coupledlinstab}) the coefficient of $\lambda$ is always positive, i.e.\ $k^2(1+d) - \gamma(f_u+g_v+f_{\widetilde u}+g_{\widetilde v}) > 0$. Hence we require that 
\begin{widetext}
\[
h^{(1)}(k^2):= dk^4 - \gamma k^2(df_u + g_v + df_{\widetilde u}+ g_{\widetilde v}) + \gamma^2(f_ug_v-f_vg_u + f_ug_{\widetilde v} - f_{\widetilde v}g_u + f_{\widetilde u}g_v - f_vg_{\widetilde u} + f_{\widetilde u}g_{\widetilde v} - f_{\widetilde v}g_{\widetilde u}) < 0
\]
\end{widetext}
holds for some wavenumber $k\ne 0$. Since we know from the linear stability conditions \eqref{eq:coupledlinstab} that the constant term is positive, i.e.\ $f_ug_v-f_vg_u + f_ug_{\widetilde v} - f_{\widetilde v}g_u + f_{\widetilde u}g_v - f_vg_{\widetilde u} + f_{\widetilde u}g_{\widetilde v} - f_{\widetilde v}g_{\widetilde u} > 0$, it follows that $df_u + g_v + df_{\widetilde u}+ g_{\widetilde v}>0$ must hold. We further need to ensure that the function $h^{(1)}$ attains a negative value for some of the wave numbers. The minimum of $h^{(1)}$ is attained at
\[
k^2_{1,\textup{m}} = \gamma\frac{df_u + g_v + df_{\widetilde u}+ g_{\widetilde v}}{2d},
\]
and the minimum value of $h^{(1)}$ is
\begin{widetext}
\[
h^{(1)}_{\textup{min}} = h^{(1)}(k_{1,\textup{m}}^2) =  \gamma^2\left[(f_ug_v-f_vg_u + f_ug_{\widetilde v} - f_{\widetilde v}g_u + f_{\widetilde u}g_v - f_vg_{\widetilde u} + f_{\widetilde u}g_{\widetilde v} - f_{\widetilde v}g_{\widetilde u}) - \frac{(df_u + g_v + df_{\widetilde u}+ g_{\widetilde v})^2}{4d}\right].
\]
\end{widetext}
The minimum value of $h^{(1)}$ is thus negative if the expression in the bracket is negative. If the first factor of \eqref{eq:kM} does not have roots with positive real part, the second factor has to have at least one root with positive real part to obtain a Turing instability. By similar reasoning as before we know from \eqref{eq:coupledlinstab} that the coefficient of $\lambda$ is again always positive: $k^2(1+d) - \gamma(f_u+g_v - f_{\widetilde u}-g_{\widetilde v}) > 0$. Hence, it is required that
\begin{widetext}
\[
h^{(2)}(k^2):= dk^4 - \gamma k^2(df_u + g_v - df_{\widetilde u} - g_{\widetilde v}) + \gamma^2(f_ug_v-f_vg_u - f_ug_{\widetilde v} + f_{\widetilde v}g_u - f_{\widetilde u}g_v + f_vg_{\widetilde u} + f_{\widetilde u}g_{\widetilde v} - f_{\widetilde v}g_{\widetilde u})<0
\]
\end{widetext}
holds for some $k\ne 0$. A necessary condition for this is $df_u + g_v - df_{\widetilde u} - g_{\widetilde v} >0$, since the constant term in $h^{(2)}$ is positive again by \eqref{eq:coupledlinstab}. To obtain a sufficient condition we have to calculate the minimum of $h^{(2)}$ as before, i.e.\
\[
k^2_{2,\textup{m}} = \gamma\frac{df_u + g_v - df_{\widetilde u} - g_{\widetilde v}}{2d}.
\]
The minimum value of $h^{(2)}$ is
\begin{widetext}
\[
h^{(2)}_{\textup{min}} = h^{(2)}(k_{2,\textup{m}}^2) =  \gamma^2\left[(f_ug_v-f_vg_u - f_ug_{\widetilde v} + f_{\widetilde v}g_u - f_{\widetilde u}g_v + f_vg_{\widetilde u} + f_{\widetilde u}g_{\widetilde v} - f_{\widetilde v}g_{\widetilde u}) - \frac{(df_u + g_v - df_{\widetilde u} - g_{\widetilde v})^2}{4d}\right].
\]
\end{widetext}

In summary, the steady state has to be linearly stable if no diffusion is present, which means that all roots of \eqref{eq:kJ} have negative real part, but instability appears when diffusion is added, which means that the polynomial in \eqref{eq:kM} has to have at least one root with positive real part. Hence for Turing instability in the coupled system \eqref{eq:coupledsys} one of the following sets of conditions has to be satisfied (\eqref{eq:Turingcondsysa} \emph{or} \eqref{eq:Turingcondsysb}):
\begin{widetext}
\begin{subequations}
\label{eq:Turingcondsys}
\begin{equation}
\label{eq:Turingcondsysa}
\begin{split}
f_u + g_v < \pm (f_{\widetilde u} + g_{\widetilde v}),\qquad f_ug_v - f_vg_u + f_{\widetilde u}g_{\widetilde v} - f_{\widetilde v}g_{\widetilde u} > \pm(f_ug_{\widetilde v} - f_{\widetilde v}g_u + f_{\widetilde u}g_v - f_vg_{\widetilde u}),\\
df_u + g_v + df_{\widetilde u}+ g_{\widetilde v}>0,\\
(df_u + g_v + df_{\widetilde u}+ g_{\widetilde v})^2 - 4d(f_ug_v-f_vg_u + f_ug_{\widetilde v} - f_{\widetilde v}g_u + f_{\widetilde u}g_v - f_vg_{\widetilde u} + f_{\widetilde u}g_{\widetilde v} - f_{\widetilde v}g_{\widetilde u}) > 0;
\end{split}
\end{equation}
\vspace{5mm}
\begin{equation}
\label{eq:Turingcondsysb}
\begin{split}
f_u + g_v < \pm (f_{\widetilde u} + g_{\widetilde v}),\qquad f_ug_v - f_vg_u + f_{\widetilde u}g_{\widetilde v} - f_{\widetilde v}g_{\widetilde u} > \pm(f_ug_{\widetilde v} - f_{\widetilde v}g_u + f_{\widetilde u}g_v - f_vg_{\widetilde u}),\\
df_u + g_v - df_{\widetilde u} - g_{\widetilde v}>0,\\ 
(df_u + g_v - df_{\widetilde u} - g_{\widetilde v})^2 -4d(f_ug_v-f_vg_u - f_ug_{\widetilde v} + f_{\widetilde v}g_u - f_{\widetilde u}g_v + f_vg_{\widetilde u} + f_{\widetilde u}g_{\widetilde v} - f_{\widetilde v}g_{\widetilde u}) >0,
\end{split}
\end{equation}
\end{subequations}
\end{widetext}
where the first line comes from the linear stability condition (hence they are the same in both cases) and the other two lines are derived from the diffusion-driven instability conditions. 

\section*{Appendix III: Cellular Models}

Here we present the details of the implementation of the cellular models presented in Figure \ref{Fig2}. We consider 1D, 2D and 3D cellular models. In all cases we solved Eqs. \eqref{eq:1}-\eqref{eq:2}, but with some terms restricted to certain subdomains as specified below. All equations were solved on the same mesh. 

\subsection*{1D Cellular Models}

We use a 1D domain, comprising $N$ subdomains of equal length (Fig.\ \ref{Fig2}b). On every subdomain the set of Eqs. \eqref{eq:1}-\eqref{eq:2} is solved. Ligand $L$ can diffuse freely in the entire domain, while receptor $R$ is restricted to each subdomain by no-flux boundary conditions. Ligand exchange between subdomains is obtained by enforcing continuous ligand profiles across the borders of the subdomains, i.e. by requiring that the ligand value $L$ on the right hand side boundary of subdomain $i$ is the same as the ligand value $L$ on the left hand side boundary of subdomain $i+1$. 

\subsection*{2D Cellular Models}

We use a 2D square domain, containing $N\times N$ equal sized subdomains of square shape. The subdomains neither intersect nor overlap (Fig. \ref{Fig2}b). The following set of PDEs is defined on this 2D domain:
\begin{eqnarray}
\hspace*{-0.4cm}  \frac{\partial R}{\partial t} &=& \Delta R+ \gamma (a-R+R^2 L) ~\text{on} ~ C\\ \label{eq:domainsC}
\hspace*{-0.4cm}  \frac{\partial L}{\partial t}  &=& d \Delta L + \gamma \begin{cases} (b-R^2 L) & ~\text{on} ~ C\\ 
 0 & ~ \text{on} ~ EC \\\end{cases}   \label{eq:domainsEC}
\end{eqnarray}
where $C$ represents the $N\times N$ array of rectangular cellular subdomains, and EC refers to the rest of the 2D domain, representing the extracellular space.

\subsection*{3D Cellular Models}

We use a 3D domain (Fig. \ref{Fig2}b), containing $N\times N\times 1$ non-overlapping spheres that are embedded into a cuboid. The following set of PDEs describes the ligand and receptor dynamics on the surface of the spheres, referred to as $C$:
\begin{eqnarray}
\hspace*{-0.4cm}  \frac{\partial R}{\partial t} &=& \Delta R+ \gamma (a-R+R^2 L) \mbox{ on } C\\ \label{eq:domainsC}
\hspace*{-0.4cm}  \frac{\partial L}{\partial t} &=& d \Delta L+ \gamma(b-R^2 L) \mbox{ on } C
\end{eqnarray}
Additionally, ligand is free to diffuse in the bulk of cuboid, referred to as $EC$:
\begin{eqnarray}
\hspace*{-0.4cm}  \frac{\partial L}{\partial t} &=& d \Delta L \mbox{ on } EC \label{eq:domainsEC}
\end{eqnarray}
The concentration of the ligand on the surface of the spheres and in the bulk of the cuboid is linked via
\begin{eqnarray}
\hspace*{-0.4cm} d \,\vec{n} \cdot \nabla L &=&\gamma(b-R^2 L) 
\end{eqnarray}
where $\vec{n}$ is the outward normal vector. The volume inside the spheres (i.e. the cell interior) is not included in the simulations because we do not consider ligand or receptor internalization.

\section*{Appendix IV: Numerical Solution of PDEs with COMSOL}
The partial differential equations were solved in COMSOL Multiphysics 4.x as described previously \cite{Vollmer:CordConferenceProceedings:2013, Germann:ProceedingsOfComsolConference2011:2011, Menshykau:ProceedingsOfComsolConference2012:2012}. COMSOL Multiphysics has previously been used to accurately solve a variety of reaction-diffusion equations which originate from chemical, biological and engineering applications \cite{Menshykau2012, Menshykau:PhysicalBiology:2013, Kotha2014, Cutress2010, Sun2006, Seymen2014, Brian2012, Adivarahan2013}. In the following we present two tests for the numerical accuracy of the solution of Turing type models obtained with COMSOL Multiphysics.

\begin{figure}[t!]
\begin{center}
\includegraphics[width=\columnwidth]{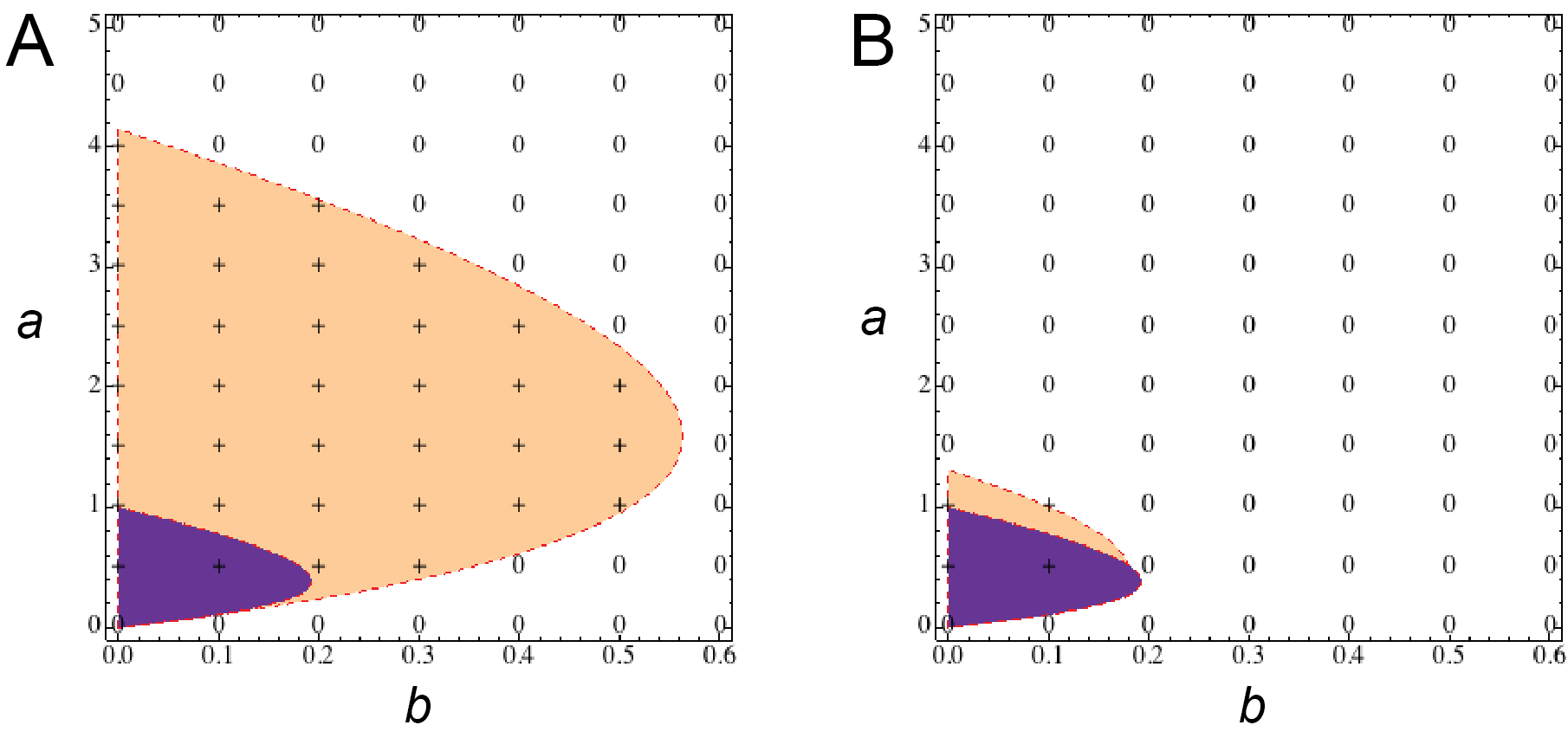}
\end{center}
\caption{
{\bf Comparison of the Turing spaces calculated numerically and those derived analytically.}  (Color online)  (A, B) The shaded regions of the parameter space indicates the area, where the linear stability analysis identifies a Turing instability (yellow, light shading) or other instabilities (navy, dark shading) for Eqs \ref{eq:1}, \ref{eq:2}  with zero-flux boundary conditions. The symbols indicate the points in the parameter space where the numerical solution of Eqs \ref{eq:1}, \ref{eq:2} with zero-flux boundary conditions yielded either pattern formation (+) or not (0). $\gamma$ was chosen sufficiently large that Turing patterns could emerge on the 1D domain.  Panel A and B differ in the relative diffusion coefficient $d$, with (A) $d=100$, and (B) $d=10$. }
\label{fig:NumTest1}
\end{figure}

\subsection{Accuracy of the Turing space}

We first test whether we obtain the same Turing space numerically and analytically. To this end, we use Eq. \ref{eq:Turingcond} as analytical condition for a Turing instability for the Turing model given by Eqs \ref{eq:1}, \ref{eq:2}. To estimate the size of the Turing space numerically, we solve Eqs \ref{eq:1}, \ref{eq:2} with COMSOL. Figure \ref{fig:NumTest1} shows that the numerical solution of Eqs \ref{eq:1}, \ref{eq:2}  in COMSOL yields pattern (+ symbols) in the part of the parameter space where the analytical criterion specifies either the classical Turing space (yellow region) or an unstable steady state both in the presence and absence of diffusion (blue region).

\subsection{Convergence of Numerical Solution}

\begin{figure}[t!]
\begin{center}
\includegraphics[width=\columnwidth]{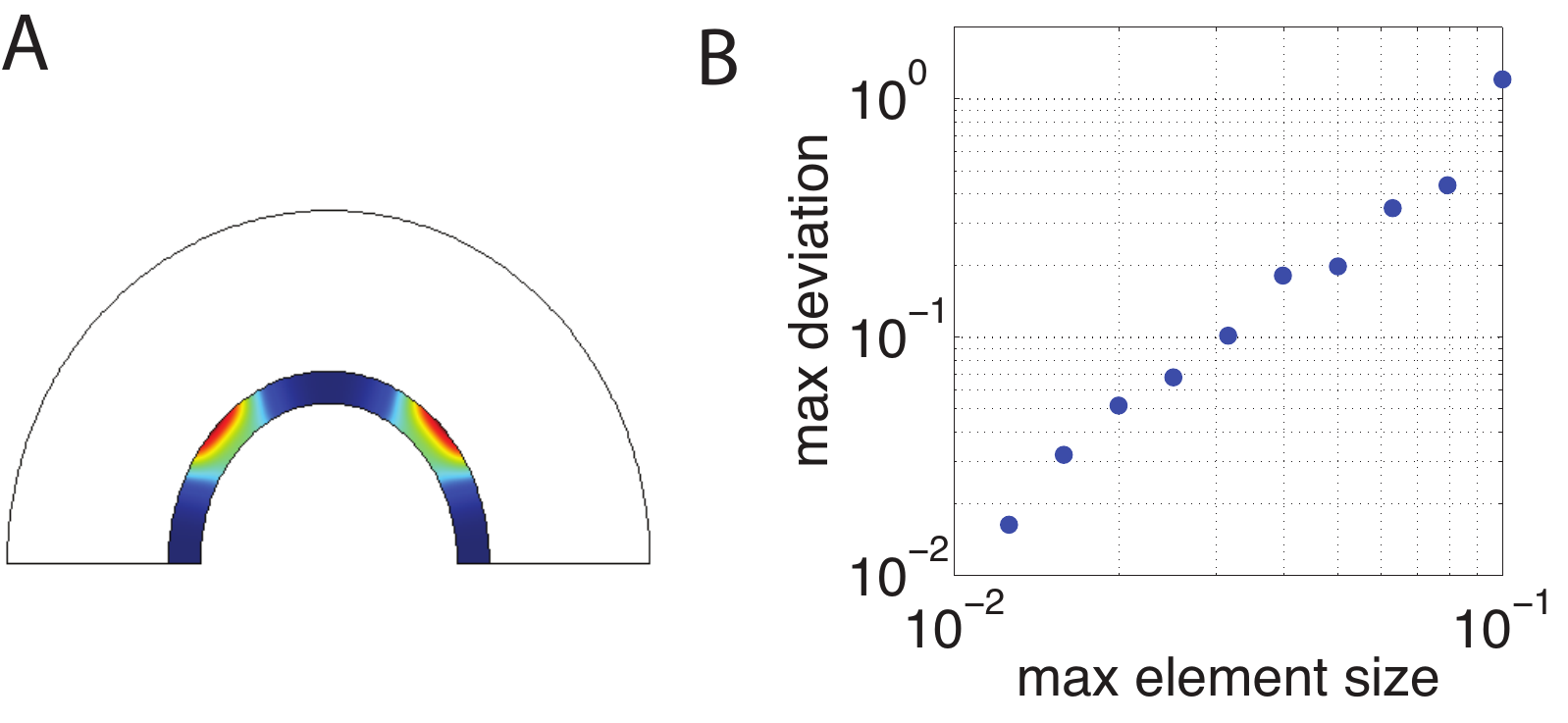}
\end{center}
\caption{
{\bf Convergence of the Numerical Solution.} (Color online)  (A) Typical pattern of receptor-ligand complexes ($R^2L$) on a domain comprising two subdomains. Ligand is produced in the upper domain, but free do diffuse on the entire domains. Receptor is produced in the lower domain and its diffusion is restricted to the lower domain. (B) The maximum deviation of the receptor-ligand complex ($R^2L$) as computed with an FEM mesh with element size equal to 0.01 from that computed at other mesh sizes.}
\label{fig:NumTest2}
\end{figure}

Here we show that the numerical solution of a ligand-receptor based Turing model on a domain comprising two layers 
converges with respect to the mesh size. We consider the model
\begin{eqnarray}
\hspace*{-0.4cm}  \frac{\partial R}{\partial t} &=& \Delta R+ \gamma (a-R+R^2 L) ~\text{on} ~T_1 \label{eq:domainsR} \\
\hspace*{-0.4cm}  \frac{\partial L}{\partial t}  &=& d \Delta L + \gamma \begin{cases} (-R^2 L) & \quad \text{on} ~T_1\\ 
 b & \quad \text{on} ~T_2 \\\end{cases}  \label{eq:domainsL}
\end{eqnarray}
where $T_1$ and $T_2$ indicate two different tissue layers. Figure \ref{fig:NumTest2}A shows the calculated distribution of the receptor-ligand complex ($R^2L$); similar patterns were obtained for a range of finite element meshes with the maximum size of the mesh size in the range from 0.01 to 0.1.  Figure \ref{fig:NumTest2}B shows that the maximum deviation in the solution decreases quadratically with respect to the maximum mesh size or equivalently decreases linearly with respect to the maximum mesh edge, as expected for FEM with first order Lagrange elements. These tests support the previous observations by others that COMSOL Multyphysics can solve Turing-type equations accurately.

\bibliographystyle{plos2009.bst}

\clearpage

\includepdf[pages=1]{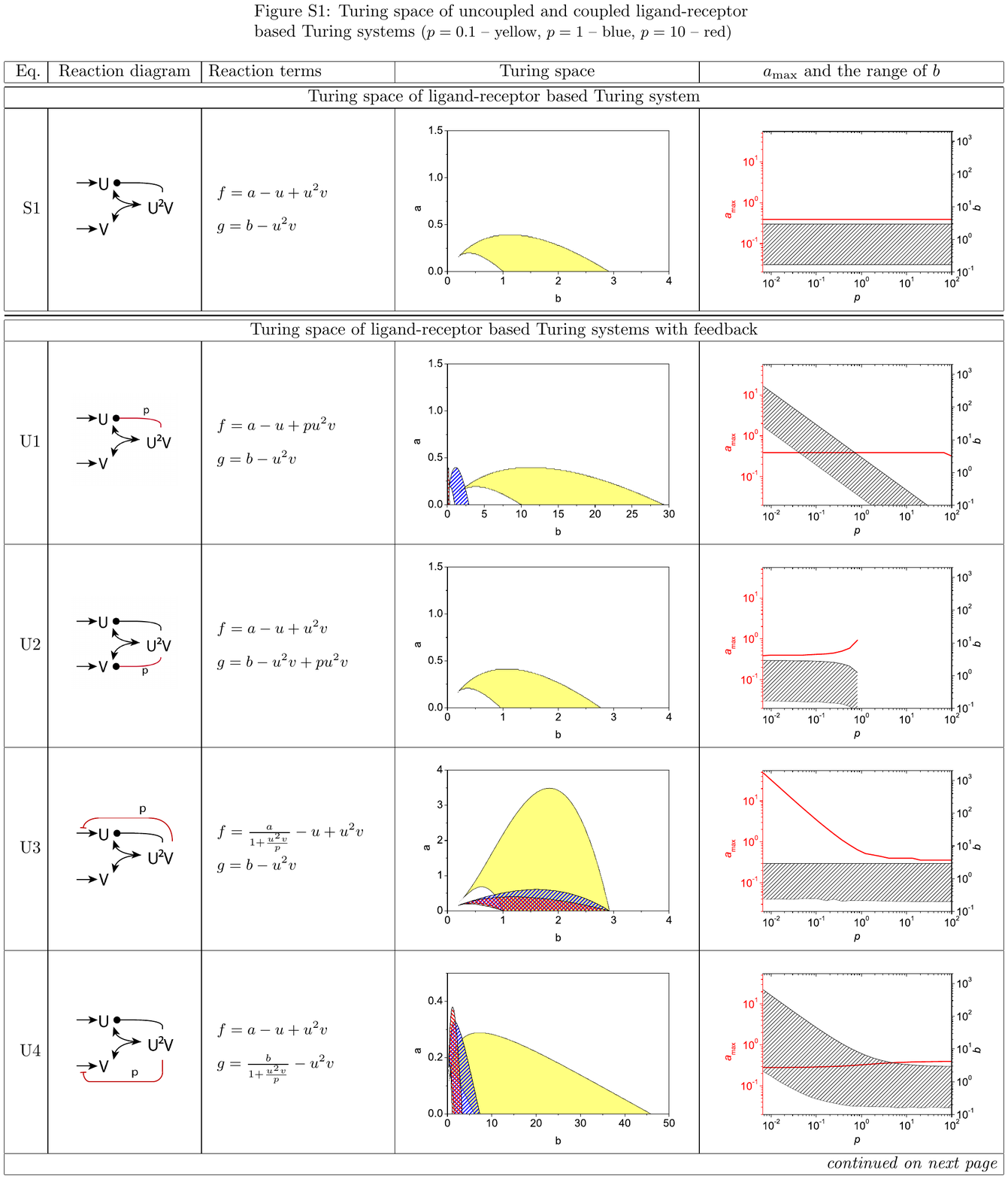}
\clearpage
\includepdf[pages=2]{Figure1_Supp.pdf}
\clearpage
\includepdf[pages=3]{Figure1_Supp.pdf}
\clearpage
\includepdf[pages=4]{Figure1_Supp.pdf}
\clearpage
\includepdf[pages=5]{Figure1_Supp.pdf}
\end{document}